\documentclass[11pt]{article}
\usepackage[english]{babel}
\usepackage[latin1]{inputenc}
\usepackage{latexsym,amsfonts,amsmath,amssymb,euscript,epsfig,dsfont,graphicx,color,float}
\usepackage[active]{srcltx}
\usepackage[T1]{fontenc}
\title{Approachability of convex sets in generalized quitting games\footnote{An extended version of the abstract has been published at the 29th edition of the annual Conference On Learning Theory, 2016.}}
\author{J\'{a}nos Flesch\footnote{Department of Quantitative Economics, Maastricht University, The Netherlands. Email: j.flesch@maastrichtuniversity.nl}, Rida Laraki\footnote{Universit\'{e} Paris-Dauphine, PSL Research University, CNRS, Lamsade, 75016 Paris, France. Also affiliated with Department of Economics, Ecole Polytechnique, France. Laraki's work
was supported by grants administered by the French National Research Agency as part of the
Investissements d'Avenir program (Idex [Grant Agreement No. ANR-11-IDEX-0003-02/Labex ECODEC
No. ANR11- LABEX-0047] and ANR-14-CE24-0007-01 CoCoRICo-CoDecreceived } and Vianney Perchet\footnote{Centre de Math\'ematiques et de Leurs Applications, ENS Cachan, France \& Criteo Labs, Paris, France. Email vianney.perchet@normalesup.org. V.\ Perchet is partially funded by the ANR grant ANR-13-JS01-0004-01  and he benefited from the support of the « FMJH Program Gaspard Monge in optimization and operation research»  and from EDF. }}

\newcommand\cX{\mathcal{X}}

\newcommand\bx{\mathbf{x}}
\newcommand\by{\mathbf{y}}

\newcommand\bI{\mathbf{I}}
\newcommand\bJ{\mathbf{J}}

\newcommand\cC{\mathcal{C}}

\newcommand\cE{\mathcal{E}}
\newcommand\cI{\mathcal{I}}
\newcommand\cJ{\mathcal{J}}
\newcommand\cM{\mathcal{M}}

\newcommand\N{\mathds{N}}
\newcommand\R{\mathds{R}}
\renewcommand\P{\mathds{P}}
\newcommand\E{\mathds{E}}

\newcommand\qed{$\hfill \blacksquare$}

\begin{document}
\maketitle
\newcounter{compteur}
 \newcounter{hypothese}
\newtheorem{proposition}[compteur]{Proposition}
\newtheorem{theorem}[compteur]{Theorem}
\newtheorem{lemma}[compteur]{Lemma}
\newtheorem{corollary}[compteur]{Corollary}
\newtheorem{hypo}[hypothese]{Assumption}
\newtheorem{definition}[compteur]{Definition}
\newtheorem{remark}[compteur]{Remark}
\newtheorem{example}[compteur]{Example}

\newcommand{\qede}{\mbox{}\hspace*{\fill}\nolinebreak \mbox{$\Diamond$}}
\newcommand{\qedr}{\mbox{}\hspace*{\fill}\nolinebreak\mbox{$\lhd$}}
\newcommand{\ep}{\varepsilon}
\newcommand{\la}{\lambda}
\newcommand{\dN}{\mathbb{N}}
\newcommand{\oz}{\overline{z}}

%\numberwithin{compteur}{section}
% \numberwithin{compteur}{subsection}

\begin{abstract}
We consider  Blackwell approachability, a very powerful and geometric tool in game theory, used for example to design strategies of the uninformed player in repeated games with incomplete information. We extend this theory to ``generalized quitting games'', a class of repeated stochastic games in which each player may have quitting actions, such as the Big-Match. We provide three simple geometric and strongly related conditions for the weak approachability of a convex target set. The first is sufficient: it guarantees that, for any fixed horizon, a player has a strategy  ensuring that the expected time-average payoff vector converges to the target set as horizon goes to infinity. The third is necessary: if it is not satisfied, the opponent can weakly exclude the target set. 

In the special case where only the approaching player can quit the game (Big-Match of type I), the three conditions are equivalent and coincide with Blackwell's condition. Consequently, we obtain a full characterization and prove that the game is weakly determined - every convex set is either weakly approachable or weakly excludable. 

In games where only the opponent can quit (Big-Match of type II), none of our conditions is both sufficient and necessary for weak approachability. We provide a  continuous time sufficient condition using techniques coming from differential games,   and show its usefulness in practice,  in the spirit of Vieille's seminal work for weak approachability.

Finally, we study uniform approachability where the strategy  should not depend on the horizon and demonstrate that, in contrast with classical Blackwell approachability for convex sets, weak approachability does not imply uniform approachability.

%More specifically, in the special case when only the decision maker can terminate the game, our two conditions are equivalent (and coincide with Blackwell's condition), which allows a full characterization of  approachable convex sets.  However, if only nature can terminate, there is a gap to which we give some insight. %Interestingly, the former condition is even sufficient for uniform approachability when only one player can quit the game.
%In big match games of type II (only the non-approaching player can quit the game), the necessary and sufficient condition is more evolved and is related to viability. 
\end{abstract}

%\begin{keywords}
%\begin{keyword}[class=AMS]
%\kwd[Primary ]{62L05}
%%\kwd[; secondary ]{62C20}
%\end{keyword}
%\begin{keyword}[class=KWD]

\section{Introduction}

We study a class of 2-player stochastic games with vector payoffs, building upon the classical models proposed by Shapley~\cite{Sha53} and Blackwell~\cite{Bla56}. \smallskip

\textbf{A finite zero-sum stochastic game} is a repeated game with perfect monitoring where the action spaces are finite and the stage payoff $g_t\in\mathbb{R}$ depends on a state-parameter that can have finitely many different values and whose evolution is controlled by both players. Shapley~\cite{Sha53} proved that such a game with the $\lambda$-discounted evaluation $\sum_{t}\lambda(1-\lambda)^{t-1}g_t$ has a value $v_\lambda$. This existence result extends easily to more general evaluations, such as $\sum_{t}\theta_t g_t$ where $\theta_t$ is the weight of stage $t$, see Laraki and Sorin~\cite{LarSor14}. For example, in the classical $T$-stage game, where $\theta_t=\tfrac{1}{T}$ for $1\leq t\leq T$, the corresponding value is denoted $v_T$. Bewley and Kohlberg~\cite{BewKoh76} proved that every stochastic game has an asymptotic value $v$, i.e. $v_\lambda$ and $v_T$ both converge to the same limit $v$, as $\lambda\rightarrow 0$ and $T\rightarrow \infty$. Mertens and Neyman~\cite{MerNey81} showed that the players can guarantee $v$ uniformly in the sense that for every $\epsilon>0$, each player has a strategy that guarantees $v$ up to $\ep$ simultaneously in every $\lambda$-discounted game for sufficiently small $\lambda$ and in every $T$-stage game for sufficiently large $T$. Such a result was earlier obtained by Blackwell and Ferguson~\cite{BlaFer68} for the game Big-Match (introduced by Gillette~\cite{Gil57}) and for the class of all absorbing games by Kohlberg~\cite{Koh74}. In fact, absorbing games are one of the very few classes where the asymptotic value has an explicit formula in terms of the one-shot game, see Laraki~\cite{Lar09}. We recall here that the asymptotic value may even fail to exist if we drop any of the following assumptions: finite action spaces, see Vigeral~\cite{Vig13}, finite state space, or perfect monitoring, see Ziliotto~\cite{Zil16}. Ergodic stochastic games, where all states are visited infinitely often almost surely regardless the actions chosen by the players, are another example.\smallskip

\textbf{A Blackwell approachability problem }is a 2-player repeated game with perfect monitoring and stage vector-payoffs $g_t\in\mathbb{R}^d$ in which player~1's objective is to enforce the convergence of $\tfrac{1}{T}\sum_{t=1}^T g_t$  to some target set $\cC \subset \mathbf{R}^d$. On the other hand, player~2 aims at preventing and his ultimate objective is to exclude this target set, i.e., to approach the complement of some $\delta$-neighborhood of it. Blackwell~\cite{Bla56} proved that the game is uniformly determined for any convex target set: either player~1 can uniformly approach $\cC$ or player~2 can uniformly exclude $\cC$. More importantly, Blackwell provided a simple geometric characterization of approachable sets from which one can easily build an optimal strategy. However, if $\cC$ is not convex, uniform determinacy fails. This led Blackwell to define a weaker version of determinacy by allowing the strategy to depend on the horizon $T$. Several years later, Vieille~\cite{Vie92} solved the problem and proved that a set is weakly approachable if and only if the value of an auxiliary differential game is zero. Weak determinacy follows from the existence of the value in differential games.\smallskip

\textbf{Combining the models of Shapley and Blackwell:} It is natural to consider stochastic games with vector payoffs and try to characterize approachable target sets in these games, notably to develop  new tools for stochastic games with incomplete information. This  challenging problem has already been tackled but, so far, only few results have been achieved. The most relevant work, by Milman~\cite{Milman}, only apply to ergodic stochastic games and no geometric characterization of approachable sets has been provided. On a different matter, it has been remarked that uniform determinacy failed to hold in stochastic games\footnote{This remark was already made by Sorin in the eighties in a small but unpublished note; its  flavor is provided in Example \ref{ex4}.}, even in variants of Big-Match.

Guided by the history of stochastic games, we tackle  the general model of stochastic games with vector payoffs by focusing, in a  first step, on the class of absorbing games, and in particular Big-Match games. Indeed,  to obtain a simple geometric characterization of approachable sets, it is helpful to consider an underlying class of games that admit an explicit characterization for the asymptotic value.

 We call  ``generalized quitting games" the subclass of absorbing games we focus on.  This terminology refers to quitting games, in which each player has exactly one quitting action and one non-quitting action. In contrast, in our case, one or both players may have none or many quitting actions.  The game is repeated until a quitting action is chosen at some stage $t^*$, in which case it enters an absorbing state that may depend on both actions at stage $t^*$. When only player~1 (resp.\ player~2) has quitting actions, the game is called Big-Match of type I (resp.\ type II). %To avoid degenerate cases, we require that player~1 (resp.\ player~2) has at least one non-quitting action too in Big-Match games of type I (resp.\ type II). \smallskip

\textbf{Main contributions:} We introduce three strongly related simple geometric conditions on a convex target set $\cC$. They are nested (the first implies the second, which implies the third) and they all have flavors of both Blackwell's condition~\cite{Bla56} and Laraki's formula in~\cite{Lar09} for the asymptotic value in absorbing games. We prove that the first condition is sufficient for player~1 to weakly approach $\cC$ and that it can be used to build an approachability strategy, even though the explicit construction is delicate and relies on a calibration technique developed notably in Perchet~\cite{Per09}. The second condition is a useful intermediate condition, but it is neither necessary nor sufficient. The third condition is proven to be necessary: indeed, if $\cC$ does not satisfy it, then $\cC$ is weakly excludable by player~2. Finally, we show that there are convex sets that are neither weakly approachable nor weakly excludable. 

We examine Big-Match games in detail. In Big-Match games of type I, our three conditions are shown to be equivalent and to coincide with Blackwell's condition. This provides a full characterization for \textit{weak} approachability and proves that this class is \textit{weakly} determined. This contrasts with the \textit{uniform} indeterminacy, see Example \ref{ex4}, where we provide a 1-dimensional counter-example. 

In Big-Match games of type II, the first two conditions are equivalent and they are proven to be necessary and sufficient for \textit{uniform} approachability. Despite this full characterization, uniform determinacy fails. For weak approachability we show that none of the three conditions is both necessary and sufficient. We also develop in some cases, an approach based on differential game, similarly to Vieille~\cite{Vie92}.

To summarize, our analysis of Big-Match games reveals that: (1) in Big-Match games of type I, a simple full characterization is available for \textit{weak} approachability;  (2) in Big-Match games of type II, a simple full characterization is available for \textit{uniform} approachability; (3) uniform determinacy fails in both types of Big-Match games; (4) weak determinacy holds for Big-Match games of type I. Weak determinacy for Big-Match games of type II remains an open problem.\smallskip

\textbf{Almost sure approachability:} In the classical Blackwell model on convex sets, and in ergodic stochastic games with vector payoffs, weak, uniform, in expectation or almost sure approachability problems are equivalent. In our case, they all differ. In this paper we focus on weak and uniform approachability in expectation, as they appear to be very interesting and challenging in generalized quitting games. We refer to Section \ref{Section-Model} and Appendix \ref{SE:ASApproach} for more details. 
\smallskip

\textbf{Related literature:}  Blackwell approachability is frequently used in the literature of repeated games. It was used first by Aumann and Maschler~\cite{AumMas95} to construct optimal strategies in zero-sum repeated games with incomplete information and perfect monitoring. Their construction has been extended by Kohlberg~\cite{Koh75} to the imperfect monitoring case. Blackwell approachability was further used by Renault and Tomala~\cite{RenTom04} to characterize the set of communication equilibria in $N$-player repeated games with imperfect monitoring; by H\"{o}rner and Lovo~\cite{HorLov09} and H\"{o}rner, Lovo and Tomala~\cite{HorLovTom11} to characterize belief-free equilibria in $N$-player repeated games with incomplete information, and by Tomala~\cite{Tom13} to characterize belief-free communication equilibria in $N$-player repeated games. Blackwell approachability has also been used to construct adaptive strategies leading to correlated equilibria (see Hart and Mas-Colell~\cite{HarMas00}), machine learning strategies minimizing regret (see 	Blackwell~\cite{Bla56a}, Abernethy, Bartlett and Hazan~\cite{AbeBarHaz11}), and calibrating algorithms in prediction problems (see Dawid~\cite{Daw85}, Foster and Vohra~\cite{FosVoh97}, Perchet~\cite{Per14jfg,Per15dgaa}). In fact, one can show that Blackwell approachability, regret-minimization and calibration are formally equivalent (see for instance Abernethy, Bartlett and Hazan~\cite{AbeBarHaz11} or Perchet~\cite{Per14jfg}). 
\smallskip

\textbf{Applications:} Classical machine learning assumes that a one stage mistake has small consequences. Our paper allows to tackle realistic situations where the total payoff can be affected by one stage decisions. One could think of clinical trials between two treatments: at some point in time one of the two must be selected and prescribed to the rest of the patients.  At a more theoretical level, as in Aumann and Maschler~\cite{AumMas95} for zero-sum repeated games with incomplete information, our paper may be a useful step towards a characterization  of the asymptotic value of absorbing games with incomplete information and determining the optimal strategy of the non--informed player. A problem for which we know existence of the asymptotic value (see Rosenberg~\cite{Ros00}), and have some explicit characterizations of the asymptotic value in $2 \times 2$ Big-Match games (see Sorin~\cite{Sor84, Sor85}).  \smallskip

\section{Model and Main Results}\label{Section-Model}
In this section, we describe the model of generalized quitting games, the problem of Blackwell approachability, and present our main results.%, introduced in Blackwell~\cite{Bla56}, that includes regret minimization as proved by ~\cite{Bla56a,AbeBarHaz11}. The model is stated directly in the more general framework of ``quitting games'', a generalization of Markov Decision Processes (see e.g.,\cite{Ber07,GopMan14,SorKedMun14}).

\paragraph{Generalized quitting games.}  We denote by  $\bI=\cI\cup\cI^\star$ the finite set of (pure) actions of player~1 and by $\bJ=\cJ\cup\cJ^\star$ the finite set of actions of player~2. The actions in $\cI$ and $\cJ$ are called non-quitting, and the actions in $\cI^\star$ and $\cJ^\star$ are called quitting. A payoff vector $g(i,j)\in \R^d$ is associated to each pair of actions $(i,j) \in \bI\times \bJ$,  and to ease notations we assume that $\|g(i,j)\|_2 \leq 1$.

The game is played at stages in $\mathbb{N}^\star$ as follows: at stage $t\in\mathbb{N}^\star$, the players choose actions simultaneously, say $i_t\in \bI$ and $j_t\in\bJ$. If only non-quitting actions have been played before stage $t$, i.e. $i_{t'}\in\cI$ and $j_{t'}\in\cJ$ for every $t'<t$, then player~1 is free to choose any action in $\bI$ and player~2 is free to choose any action in $\bJ$. However, if a quitting action was played by either player at a stage prior to stage $t$, i.e. $i_{t'}\in\cI^\star$ or $j_{t'}\in\cJ^\star$ for some $t'<t$, then the players are obliged to take $i_t=i_{t-1}$ and $j_t=j_{t-1}$. Another equivalent way to model this setup is to assume that, as soon as a quitting action is played, the game absorbs in a state where the payoff is constant.

When a player plays a quitting action and neither player has played a quitting action before, we say that this player quits and that play absorbs.

\paragraph{Mixed actions.} A mixed action for a player is a probability distribution over his (pure) actions. We will denote mixed actions of player~1 by $\bx \in \Delta(\cI\cup\cI^\star)$, $x \in \Delta(\cI)$, $x^\star \in \Delta(\cI^\star)$. Thus, a bold letter stands for a mixed action over the full set of actions $\bI$, a regular letter for a mixed action restricted to non-quitting actions in $\cI$ and a letter with an asterisk for a mixed action over the set of quitting actions in $\cI^\star$. Similarly, we denote mixed actions of player~2 by $\by \in \Delta(\cJ\cup\cJ^\star)$, $y \in \Delta(\cI)$ and $y^\star \in \Delta(\cI^\star)$. 

To introduce our conditions for a convex set to be approachable, it will be helpful to consider finite nonnegative measures on $\bI$ and $\bJ$ instead of probability distributions. We shall denote them by $\alpha \in \cM(\bI)$ for player~1 and by $\beta \in \cM(\bJ)$ for player~2.

The payoff mapping $g$ is extended as usual multi-linearly to the set of mixed actions $\Delta(\bI)$ and $\Delta(\bJ)$ and, more generally, to the set of measures $\cM(\bI)$ and $\cM(\bJ)$.

We also introduce the ``measure'' or ``probability of absorption'' and the ``expected absorption payoff'' (which is not the expected payoff conditional to absorption), defined  respectively by
$$ p^\star(\alpha,\beta)=
\sum_{(i,j)\hspace{0.05cm}\in\hspace{0.05cm} (\cI^\star\times \bJ)\cup (\bI\times \cJ^\star)} \alpha_{i}  \beta_{j} \quad \text{ and } \quad g^\star(\alpha,\beta)=\sum_{(i,j)\hspace{0.05cm}\in\hspace{0.05cm} (\cI^\star\times \bJ)\cup (\bI\times \cJ^\star)} \alpha_{i}\beta_{j}g(i,j).$$

\paragraph{Strategies.} In our model of generalized quitting games, histories are defined as long as no quitting action is played. Thus, the set of histories $H$ is the set of finite sequences in $\cI\times \cJ$ (that is, $H=\cup_{t=1}^{\infty} (\cI\times \cJ)^{t-1}$). A strategy for player~1 is a mapping $\sigma:H\rightarrow \Delta(\bI)$, and a strategy for player~2 is a mapping $\tau:H\rightarrow \Delta(\bJ)$. 

\paragraph{Specific subclasses of games: Big-Match games.} We shall consider two subclasses of generalized quitting games, in which only one of the players can quit. 

Following the nomenclature of Gillette~\cite{Gil57}, a generalized quitting game is called a Big-Match game of type I if player~1 has at least one non-quitting action -- to avoid degenerate cases --  and at least one quitting action, but player~2 has no quitting action, i.e. $\cI\neq \emptyset$, $\cI^*\neq \emptyset$ and $\cJ^\star=\emptyset$. 

A generalized quitting game is called a Big-Match game of type II if player~2 has at least one non-quitting action and at least one quitting action, but player~1 has no quitting action, i.e. $\cJ\neq \emptyset$, $\cJ^*\neq \emptyset$ and $\cI^\star=\emptyset$.
 
\paragraph{Objectives.} In short, the objective of player~1 is to construct a strategy $\sigma$ such that, for any strategy $\tau$ of player~2, the expected average payoff $\E_{\sigma,\tau} \frac{1}{T}\sum_{t=1}^T g(i_t,j_t)$ is close to some exogenously given convex set $\cC \subset \R^d$, called the ``target set''. Instead of the Cesaro average, we can also consider the expected discounted payoff $\E_{\sigma,\tau} \sum_{t=1}^\infty \lambda(1-\lambda)^{t-1}g(i_t,j_t)$ or even a general payoff evaluation $\E_{\sigma,\tau} \sum_{t=1}^\infty \theta_tg(i_t,j_t)$, where $\theta_t \in [0,1]$ and $\sum_{t=1}^{\infty} \theta_t =1$, with the interpretation that $\theta_t$ is the weight of stage $t$.

We emphasize here that we focus on the distance of the expected average payoff to $\cC$ (and not on the expected distance of the average payoff to $\cC$, corresponding to almost sure convergence, see e.g.\ Milman~\cite{Milman}), as it is might be more traditional and even challenging in stochastic games. Indeed, consider the toy game where player~1 has only two actions, both absorbing, and they give payoffs $-1$ and $1$ respectively. In this game, $\{0\}$ is obviously not approachable in the almost sure sense, but is easily approachable in the expected sense by playing each action with probability $\frac{1}{2}$ at the first stage. We still quickly investigate almost sure approachability in Appendix \ref{SE:ASApproach}.
\medskip

We can distinguish at least two different concepts of approachability, that we respectively call %\footnote{We purposely depart from the vocabulary of ~\cite{Bla56} or ~\cite{Vie92}, which corresponds to the concept of asymptotic value in the literature of repeated games.} 
 uniform approachability and weak approachability.

\medskip

Specifically, we say that a convex set $\cC \subset \R^d$ is uniformly approachable by player~1, if for every $\varepsilon >0$ player~1 has a strategy  such that after some stage $T_\varepsilon \in \N$, the expected average payoff is $\varepsilon$-close to $\cC$, against any strategy of player~2. Stated with quantifiers
$$ \cC \text{ is uniformly app. } \iff \forall \varepsilon >0, \exists \sigma, \exists T_\varepsilon \in \N, \forall T\geq T_\varepsilon, \forall \tau, d_\cC\Big( \E_{\sigma,\tau} \frac{1}{T}\sum_{t=1}^Tg(i_t,j_t) \Big) \leq \varepsilon.
$$
Reciprocally, a convex set $\cC\subset \R^d$ is uniformly excludable by player~2 if she can uniformly approach the complement of some $\delta$ neighborhood of $\cC$, for some fixed $\delta>0$.

%Usually, approachability is stated in terms of the expected distance of the average payoff to $\cC$ and not the distance of the expected payoff. But as argued above, in our games, the latter might be a too strong requirement to satisfy and moreover the problem of characterization is more easy to solve (see footnote ), and hence we focus on the distance of the expected payoff. 

A similar definition holds for general evaluations  induced by a sequence of weights $\theta=(\theta_t)_{t\in\mathbb{N}}$ such that $\sum_{t=1}^\infty \theta_t =1$ and $\theta_t \geq 0$ for all $t\in\mathbb{N}$. For every $\varepsilon>0$, there must exist a threshold $\theta_\varepsilon$  so that if the sequence $\theta=(\theta_t)_{t \in \N}$ satisfies $\|\theta\|_2=\sqrt{\sum_t \theta_t^2} \leq \theta_\varepsilon$ then the $\theta$-evaluation of the payoffs is within distance $\varepsilon$ of $\cC$. We emphasize here that the Cesaro average corresponds to $\theta_t = 1/N$ for $t \in [N]=\{1,\ldots,N\}$ while the discounted evaluation, with discount factor $\lambda \in (0,1]$ corresponds to $\theta_t = \lambda(1-\lambda)^{t-1}$ for all $t \geq 1$. We then denote the accumulated $\theta$-weighted average payoff up to stage $t \in \N \cup \{\infty\}$ as $$\overline{g}_t^\theta = \sum_{s=1}^t \theta_sg(i_s,j_s), \quad \overline{g}_t^N = \frac{1}{N} \sum_{s=1}^{t \wedge N} g(i_s,j_s), \quad \text{and} \quad \overline{g}_t^\lambda = \sum_{s=1}^t \lambda(1-\lambda)^{s-1}(i_s,j_s) .$$

\medskip

We now  focus on our main objective, weak approachability. We say that a convex set $\cC \subset \R^d$ is weakly approachable by player~1, if for every $\varepsilon >0$, if the horizon of the game is sufficiently large and known, player~1 has a strategy  such that the expected payoff is $\varepsilon$-close to $\cC$, against any strategy of player~2. Stated with quantifiers
$$ \cC \text{ is  weakly app. } \iff \forall \varepsilon >0,  \exists T_\varepsilon \in \N, \forall T\geq T_\varepsilon,\exists \sigma_T, \forall \tau, d_\cC\Big( \E_{\sigma_T,\tau} \frac{1}{T}\sum_{t=1}^Tg(i_t,j_t) \Big) \leq \varepsilon
$$
Reciprocally, a convex set $\cC\subset \R^d$ is weakly excludable by player~2 if she can weakly approach the complement of some $\delta$ neighborhood of $\cC$. This definition of weak approachability may be extended, just as above, for general evaluation, where the strategy of player~1 depends on $\theta=(\theta_t)_{t \in \N}$.

\medskip

Observe that we can assume without loss of generality that the target set $\cC$ is closed, because approaching a set or its closure are two equivalent problems.

We emphasize that, without an irreversible Markov chain structure,  uniform approachability will be equivalent to weak approachability, because of the doubling trick, when the target set is convex. However, as we shall see, it is no longer the case in generalized quitting games.  

\paragraph{Reminder on approachability in classical repeated games.} Blackwell~\cite{Bla56} proved that in classical repeated games (i.e., when $\cI^\star=\cJ^\star=\emptyset$) there is a simple geometric necessary and sufficient condition under which a convex set is (uniformly and weakly) approachable. It reads as follows
\begin{align*} \cC \text{ is uniformly/weakly app. } &\iff \forall y \in \Delta(\cJ), \exists x \in \Delta(\cI), g(x,y)\in \cC\\
& \iff \max_{y \in \Delta(\cJ)} \min_{x \in \Delta(\cI)} d_\cC(g(x,y)) =0.
\end{align*}
This immediately entails that  a convex set is either weakly approachable or weakly excludable. 
\paragraph{Approachability conditions.} We aim at providing a similar geometric condition ensuring that a convex set $\cC$ is weakly approachable (or weakly excludable). Inspired by a recent formula, obtained in Laraki~\cite{Lar09}, which characterizes the asymptotic value by making use of perturbations of mixed actions with measures, we introduce the following three conditions.

The strongest of the three conditions is:
\begin{equation}\label{EQ:Sufficient}
\max_{\by \in \Delta(\bJ)} \min_{\bx \in \Delta(\bI)}\inf_{\alpha \in \cM(\bI)}\sup_{\beta \in \cM(\bJ)} d_\cC \Big(\frac{g(\bx,\by)+g^\star(\alpha,\by)+g^\star(\bx,\beta)}{1+p^\star(\alpha,\by)+p^\star(\bx,\beta)} \Big) =0.
\end{equation}

The next condition will be shown to be a useful intermediate condition:
\begin{equation}\label{EQ:non-Necessary}
\max_{\by \in \Delta(\bJ)} \min_{\bx \in \Delta(\bI)}\sup_{\beta \in \cM(\bJ)}\inf_{\alpha \in \cM(\bI)} d_\cC \Big(\frac{g(\bx,\by)+g^\star(\alpha,\by)+g^\star(\bx,\beta)}{1+p^\star(\alpha,\by)+p^\star(\bx,\beta)} \Big) =0.
\end{equation}

Finally, the weakest of the three conditions is:
\begin{equation}\label{EQ:Necessary}
\max_{\by \in \Delta(\bJ)}\sup_{\beta \in \cM(\bJ)}  \min_{\bx \in \Delta(\bI)}\inf_{\alpha \in \cM(\bI)}d_\cC \Big(\frac{g(\bx,\by)+g^\star(\alpha,\by)+g^\star(\bx,\beta)}{1+p^\star(\alpha,\by)+p^\star(\bx,\beta)} \Big) =0.
\end{equation}

%\begin{itemize}
%\item[] \textsl{Sufficient condition}
%\begin{equation}\label{EQ:Sufficient}
%\max_{\by \in \Delta(\bJ)} \min_{\bx \in \Delta(\bI)}\inf_{\alpha \in \cM(\bI)}\sup_{\beta \in \cM(\bJ)} d_\cC \Big(\frac{g(\bx,\by)+g^\star(\alpha,\by)+g^\star(\bx,\beta)}{1+p^\star(\alpha,\by)+p^\star(\bx,\beta)} \Big) =0
%\end{equation}
%\item[] \textsl{Non-necessary, non-sufficient  condition}
%\begin{equation}\label{EQ:non-Necessary}
%\max_{\by \in \Delta(\bJ)} \min_{\bx \in \Delta(\bI)}\sup_{\beta \in \cM(\bJ)}\inf_{\alpha \in \cM(\bI)} d_\cC \Big(\frac{g(\bx,\by)+g^\star(\alpha,\by)+g^\star(\bx,\beta)}{1+p^\star(\alpha,\by)+p^\star(\bx,\beta)} \Big) =0
%\end{equation}
%\item[] \textsl{Necessary  condition}:
%\begin{equation}\label{EQ:Necessary}
%\max_{\by \in \Delta(\bJ)}\sup_{\beta \in \cM(\bJ)}  \min_{\bx \in \Delta(\bI)}\inf_{\alpha \in \cM(\bI)}d_\cC \Big(\frac{g(\bx,\by)+g^\star(\alpha,\by)+g^\star(\bx,\beta)}{1+p^\star(\alpha,\by)+p^\star(\bx,\beta)} \Big) =0
%\end{equation}
%\end{itemize}

We emphasize here that, in the above conditions, the maxima and minima are indeed attained since the mapping $(\bx,\alpha,\by,\beta) \mapsto \frac{g(\bx,\by)+g^\star(\alpha,\by)+g^\star(\bx,\beta)}{1+p^\star(\alpha,\by)+p^\star(\bx,\beta)}$ is Lipchitz. %Clearly, Condition \eqref{EQ:Sufficient} is the strongest, Condition \eqref{EQ:Necessary} is the weakest among these conditions and Condition \eqref{EQ:non-Necessary} lies in-between them.\footnote{More precisely, if a set satisfies condition \eqref{EQ:Sufficient}, it satisfies all the two others, etc.}

\paragraph{Main results.}
We can already state our main results, which we will prove throughout the paper. In these results, approachability always refers to player~1 whereas excludability always refers to player~2.

\begin{theorem}[Weak Approachability]\label{TH: Results} Let $\cC \subset \R^d$ be a convex set.
\begin{description}
\item \textsc{Sufficiency:} If  Condition \eqref{EQ:Sufficient} is satisfied, then $\cC$ is weakly approachable.
\item  \textsc{Necessity:} If $\cC$ is weakly approachable, then Condition \eqref{EQ:Necessary} is satisfied.
%\item  \textsc{Necessity-bis:} 
Indeed, if Condition \eqref{EQ:Necessary} is not satisfied, then $\cC$ is weakly excludable.
\end{description}
\end{theorem}

The theorem above is proven in Section 3.

In the special class of Big-Match games, our findings for weak approachability are summarized by the next proposition.
\begin{proposition}[Weak Approachability in Big-Match Games]\label{MProp2}\ 
\begin{description}
\item  \textsc{In Big-Match games of type I:} Conditions \eqref{EQ:Sufficient}, \eqref{EQ:non-Necessary}  and \eqref{EQ:Necessary} coincide with the classical Blackwell condition, i.e., 
$$\cC \text{ is  weakly approachable } \ \iff\  \forall \by \in \Delta(\bJ), \exists \bx \in \Delta(\bI), g(\bx,\by)\in\cC.$$ Consequently, in this class, weak determinacy holds: a convex set is either weakly approachable or weakly excludable.
%\item[ii)] In Big-match game of type II,  when player~1 cannot quit ($\cI^\star=\emptyset$), a closed and convex set  $\cC \subset \R^d$  is weak-approachable if it is uniformly approachable.
%\item[ii)] 
\item  \textsc{In Big-Match games of type II:} Conditions \eqref{EQ:Sufficient}  and  \eqref{EQ:non-Necessary}  coincide and generally differ with Condition \eqref{EQ:Necessary}. Moreover, none of the conditions is both sufficient and necessary for weak approachability.
\end{description}
\end{proposition}
%Moreover, condition \eqref{EQ:non-Necessary} is not necessary nor sufficient for weak approachability. 
%and there are convex sets that are neither weakly approachable nor weakly excludable.

In the proposition above, the first part of the claim on Big-Match games of type I follows from Lemma \ref{LM:EqSufTypeI}. The second part is then a direct consequence of Theorem \ref{TH: Results}. Indeed, suppose that a convex set $\cC$ is not weakly approachable. Then by Theorem \ref{TH: Results}, $\cC$ does not satisfy Condition \eqref{EQ:Sufficient}. Since Conditions \eqref{EQ:Sufficient} and \eqref{EQ:Necessary} coincide, $\cC$ does not satisfy Condition \eqref{EQ:Necessary} either. Hence, by Theorem \ref{TH: Results} once again, $\cC$ is weakly excludable.

The claim on Big-Match games of type II follows from Lemma \ref{LM:eqBigMatch2} and by Examples \ref{ex5} and \ref{ex6}.

For uniform approachability we obtain the following results.
\begin{proposition}[Uniform Approachability in Big-Match Games]\ 
\begin{description}
\item  \textsc{In Big-Match games of either type:} There are convex sets which are weakly approachable but not uniformly approachable; and so they are neither uniformly approachable nor uniformly excludable.
\item  \textsc{In Big-Match games of type II:} 
Condition \eqref{EQ:Sufficient} (and hence Condition \eqref{EQ:non-Necessary}) is necessary and sufficient for uniform approachability. 
\end{description}
\end{proposition}

The first claim is shown by Examples \ref{ex4} and \ref{ex5}. The second claim on Big-Match games of type II follows from Proposition \ref{Prop:UnifType2}.\bigskip

Notice that, in the results above, approachability conditions for Big-Match games of types I and II are drastically different. The necessary and sufficient weak approachability condition takes a simple form for type I, but not for type II, the situation being completely reversed for uniform approachability. The second consequence is that determinacy of convex sets is very specific to the original model of Blackwell~\cite{Bla56}. We remark that determinacy also fails in the standard model of Blackwell, if player~1 has an imperfect observation on past actions of player~2,  as proved by Perchet~\cite{Per11jota} by providing an example of a convex set that  is neither approachable nor excludable.

\paragraph{Outline of the Paper.}
The remaining of the paper is organized as follows. In Section 3, we prove Theorem \ref{TH: Results}. In Section 4 we compare the notions of weak and uniform approachability, with a focus on Big-Match games. In Section 5, we present several examples. Additional results and examples can be found in the Appendices.
%
%Then, we will consider more specifically Big-match games. We will show that Conditions~\eqref{EQ:Sufficient} and \eqref{EQ:Necessary} coincide for Big-Match games of type I (thus closing the question in those games), and we will provide examples to explain why uniform and weak approachability are two different concepts. We will  also prove that Condition \eqref{EQ:non-Necessary}, which is stronger than \eqref{EQ:Necessary} and weaker than \eqref{EQ:Sufficient}, is neither sufficient nor necessary for weak approachability in Big-match games of type II and give some insight, in the line of Vieille's~\cite{Vie92} characterization of weak approachability for non convex sets via differential games and controlability.

\section{Necessary and Sufficient Conditions for Weak Approachabilty}

In this section, we prove Theorem \ref{TH: Results}. First we shall prove that, assuming the sufficiency of Condition~\eqref{EQ:Sufficient} for weak approachability, Condition \eqref{EQ:Necessary} is indeed necessary. Then, we show that Condition~\eqref{EQ:Sufficient} ensures weak approachability.

\subsection{If Condition \eqref{EQ:Sufficient} is Sufficient, then Condition \eqref{EQ:Necessary} is Necessary}\label{suff-implies-ness}
As claimed, we will prove later  in Proposition \ref{proplabel} that Condition \eqref{EQ:Sufficient} is sufficient for the weak approachability of convex sets in generalized quitting games. The purpose of this section is to demonstrate that this entails the necessity of Condition \eqref{EQ:Necessary} by switching the role of players 1 and 2.

\begin{proposition}
Assume that Condition \eqref{EQ:Sufficient} is sufficient for the weak approachability of convex sets. Then Condition \eqref{EQ:Necessary} is necessary: If a convex set $\cC$ does not satisfy Condition \eqref{EQ:Necessary}, then $\cC$ is weakly excludable by player~2.
\end{proposition}
\textbf{Proof.}
As Condition \eqref{EQ:Necessary} is not satisfied for $\cC$, there exists $\delta >0$ such that
$$\max_{\by \in \Delta(\bJ)}\sup_{\beta \in \cM(\bJ)}  \min_{\bx \in \Delta(\bI)}\inf_{\alpha \in \cM(\bI)}d_\cC \Big(\frac{g(\bx,\by)+g^\star(\alpha,\by)+g^\star(\bx,\beta)}{1+p^\star(\alpha,\by)+p^\star(\bx,\beta)} \Big) \geq \delta.$$
Choose some $\by_0$ and $\beta_0$ that realize the supremum up to $\delta/2$. It is not difficult to see that

$$\Big\{\frac{g(\bx,\by_0)+g^\star(\alpha,\by_0)+g^\star(\bx,\beta_0)}{1+p^\star(\alpha,\by_0)+p^\star(\bx,\beta_0)}\ ;\ \bx \in  \Delta(\bI), \alpha \in \cM(\bI) \Big\}$$ is a bounded convex   set that is $\delta/2$ away from $\cC$ and whose closure is denoted by $\cE$.

To prove the convexity of the above set, suppose that $$z=\sum_i \lambda_i \frac{g(\bx_i,\by_0)+g^\star(\alpha_i,\by_0)+g^\star(\bx_i,\beta_0)}{1+p^\star(\alpha_i,\by_0)+p^\star(\bx_i,\beta_0)}, \ \text{ with } \lambda_i \geq 0 \ \text{ and  } \sum_i \lambda_i=1\ . $$ Taking $\theta_i= \big(1+p^\star(\alpha_i,\by_0)+p^\star(\bx_i,\beta_0)\big)^{-1}$, we obtain that $$z=\frac{g(\bx,\by_0)+g^\star(\alpha,\by_0)+g^\star(\bx,\beta_0)}{1+p^\star(\alpha,\by_0)+p^\star(\bx,\beta_0)} \ \text{ with } \bx=\frac{\sum_i \lambda_i \theta _i \bx_i}{\sum_i \lambda_i \theta _i} \ \text{ and } \alpha=\frac{\sum_i \lambda_i \theta _i \alpha_i}{\sum_i \lambda_i \theta _i}\ . $$

Thus we have proved that
$$ \max_{\bx \in \Delta(\bI)}\min_{\by \in \Delta(\bJ)}\inf_{\beta \in \cM(\bJ)}\max_{\alpha \in \cM(\bI)}d_\cE \Big(\frac{g(\bx,\by)+g^\star(\alpha,\by)+g^\star(\bx,\beta)}{1+p^\star(\alpha,\by)+p^\star(\bx,\beta)} \Big) =0,$$
and that $\cE$ satisfies Condition \eqref{EQ:Sufficient}, but stated from the point of view of player~2. Therefore, by assumption, player~2 can weakly approach $\cE$, and hence she can weakly approach the complement of the $\delta/2$-neighborhood of $\cC$. This means that $\cC$ is weakly excludable by player~2 (and in particular not weakly approachable by player~1), as desired. \qed

\subsection{Condition \eqref{EQ:Sufficient} is Sufficient}
 
 We prove in this section that Condition \eqref{EQ:Sufficient} is sufficient for the weak approachability of a convex set $\cC \subset \R^d$. Assuming that the target set satisfies Condition \eqref{EQ:Sufficient}, the construction of the approachability strategy will be based on a calibrated algorithm, as introduced by  Dawid~\cite{Daw85}. Similar ideas can be found in the online learning literature (see, e.g., Foster and Vohra~\cite{FosVoh97}, Perchet~\cite{Per09}, and Bernstein, Mannor and Shimkin~\cite{BerManShi13}), where Blackwell approachability and calibration now play an increasingly important role (as evidenced by Abernethy, Bartlett and Hazan~\cite{AbeBarHaz11}, Mannor and Perchet~\cite{ManPer13}, Perchet~\cite{Per11jmlr}, Rakhlin, Sridharan and Tewari~\cite{RakSriTew11}, and Foster, Rakhlin, Sridharan and Tewari~\cite{FosRakSri11}).
  
For the sake of clarity, we divide this section in several parts. First, we introduce the auxiliary calibration tool, then we prove the sufficiency condition in Big-Match games of types II and I, and finally we show how the main idea generalizes.
\subsubsection{An auxiliary tool: Calibration}

In this subsection, we adapt a result of Mannor, Perchet and Stoltz~\cite{ManPerSto14jmlr} on calibration to the setup with the general payoff evaluation (thus not necessarily with Cesaro averages). 

We recall that calibration is the following sequential decision problem.  Consider a non-empty and finite set $\Omega$, a finite $\varepsilon$-grid of the set of probability distributions on $\Omega$ denoted by $\{ p_k \in \Delta(\Omega), k \in [K] \}$ where $[K]=\{1,\ldots,K\}$ for $K \in  \N$, and a sequence of weights $\{\theta_t  \in \R_+\}_{t \in \N}$. At each stage $t$, Nature chooses a state $\omega_t \in \Omega$ and the decision maker predicts it by choosing simultaneously a point of the grid $p_t\in\Delta(\Omega)$. Once $p_t$ is chosen, the state $\omega_t$ and the weight $\theta_t$ are revealed (we emphasize that the sequence $\theta$ is not necessarily known in advance by the decision maker). 

We denote by $\N_t[k]= \{ s \leq t\ \text{s.t.} \ p_s=p_k\}$ the set of stages $s \leq t$ where $p_k$ was predicted and by 
\[ \bar{\omega}_t[k]=\frac{\sum_{s \in \N_t[k]} \theta_s \delta_{\omega_s}}{\sum_{s \in \N_t[k]} \theta_s} \in \Delta(\Omega)\]
the empirical weighted distribution of the state on $\N_t[k]$. 

In that setting, we say that an algorithm of the decision maker is calibrated if
\[ \limsup_{t \to \infty} \max_{k \in [K]} \frac{\sum_{s \in \N_t[k]} \theta_s}{ \sum_{s \leq t} \theta_s}\Big( \left\|p_k-\bar{\omega}_t[k] \right\|-\varepsilon\Big)_+=0
\]
almost surely.

\begin{lemma}\label{LM:Calib}
The decision maker has a calibrated algorithm such that, for all $t \in \N$,
\[ \E\left[ \max_{k \in [K]}\Big(\sum_{s \in \N_t[k]} \theta_s\Big) \Big( \left\|p_k-\bar{\omega}_t[k] \right\|-\varepsilon\Big)_+ \right] \leq  \sqrt{8|\Omega|\sum_{s \leq t} \theta^2_s}.
\]
\end{lemma}
\textbf{Proof.} The proof is almost identical to the one in Mannor, Perchet and Stoltz~\cite{ManPerSto14jmlr}, Appendix A, thus is omitted. \qed

\medskip

We mention here that  the construction of a calibrated algorithm is actually often based on the construction of an approachability strategy, as in Foster~\cite{Fos99}, and Perchet~\cite{Per14jfg,Per15dgaa}.

\subsubsection{Condition \eqref{EQ:Sufficient} is sufficient in Big-Match games of type II}
We first focus on Big-match games of type II, where only player~2 can quit. The following lemma exhibits a useful equivalence between some of the conditions. %, where only player~2 can quit. %In those games,   Condition \eqref{EQ:Sufficient} can be rewritten in a more convenient, but less compact, way.

\begin{lemma}\label{LM:eqBigMatch2}
In Big-Match games of type II,  Condition \eqref{EQ:Sufficient} and Condition \eqref{EQ:non-Necessary} are equivalent, and they are further equivalent to
\begin{equation}\label{suff-BMII} \forall y \in \Delta(\cJ), \exists x \in \Delta(\cI), g(x,y) \in \cC \ \mathrm{ and } \ g(x,j^\star) \in \cC,  \forall j^\star \in \cJ^\star.
\end{equation}
\end{lemma}
A consequence of \eqref{suff-BMII} is that if player~2, at every stage, either plays  a non-quitting action i.i.d.\ accordingly to $y \in \Delta(\cJ)$ or decides to quit, then player~1 can approach $\cC$ by playing i.i.d.\ accordingly to $x \in \Delta(\cI)$. The sufficiency of this condition means that it is not more complicated to approach $\cC \subset \R^d$ against an opponent than against an i.i.d.\ process that could eventually quit at some (unknown) time.

\textbf{Proof:} We already know that Condition \eqref{EQ:Sufficient} implies Condition \eqref{EQ:non-Necessary}. 

Now we prove that Condition \eqref{EQ:non-Necessary} implies \eqref{suff-BMII}. So, assume that $\cC$ satisfies Condition \eqref{EQ:non-Necessary}. Since $p^\star(\alpha,y)=0$ for all $\alpha\in\cM(\bI)$ and $y\in\Delta(\cJ)$, Condition \eqref{EQ:non-Necessary} implies
\[ \forall y \in \Delta(\cJ), \exists x \in \Delta(\cI), \forall \beta \in \cM(\bJ), \frac{g(x,y)+g^\star(x,\beta)}{1+p^\star(x,\beta)} \in \cC.
\] 
Now \eqref{suff-BMII} follows by taking $\beta=0$ and respectively taking $\beta_c= c\cdot\delta_{j^\star}$ with $c$ tending to infinity.
%\[ \forall y \in \Delta(\cJ), \exists x \in \Delta(\cI), g(x,y) \in \cC \ \mathrm{ and } \ g(x,j^\star) \in \cC,  \forall j^\star \in \cJ^\star.\]

Finally, we prove that \eqref{suff-BMII} implies Condition \eqref{EQ:Sufficient}. So, assume that $\cC$ satisfies \eqref{suff-BMII}. Let $\by\in\Delta(\bJ)$.  Decompose it as $\by=\gamma y + (1-\gamma)y^\star$, where $y\in\Delta(\cJ)$, $y^\star\in\Delta(\cJ^\star)$ and $\gamma\in[0,1]$. For this $y$, let $x \in \Delta(\cI)$ be given by \eqref{suff-BMII}. Then %it holds that
%\[ g(x,\by)=\gamma g(x,y)+(1-\gamma) \sum_{j^\star \in \cJ^\star} y^\star_{j^\star} g(x,j^\star) \in \cC\]
%Since by definition of $g^\star$, we have $g^\star(x,\beta)=\sum_{j^\star \in \cJ^\star} \beta^\star_{j^\star} g(x,j^\star)$, we get that
\[ \frac{g(x,\by)+g^\star(x,\beta)}{1+p^\star(x,\beta)}=\frac{\gamma g(x,y)+(1-\gamma) \sum_{j^\star \in \cJ^\star} y^\star_{j^\star} g(x,j^\star)+\sum_{j^\star \in \cJ^\star} \beta_{j^\star} g(x,j^\star)}{\gamma+(1-\gamma)\sum_{j^\star \in \cJ^\star} y^\star_{j^\star} +\sum_{j^\star \in \cJ^\star} \beta_{j^\star}}\in \cC,\]
because  all involved payoffs, $g(x,y)$ and $g(x,j^\star)$, belong to the convex set $\cC$. Since we can choose $\alpha=0$, we have shown that Condition \eqref{EQ:Sufficient} holds, as desired. \qed

\begin{proposition}\label{lem-BMI-suff} In Big-Match games of type II, a convex set $\cC$ is (weakly or uniformly) approachable by player~1 if Condition \eqref{EQ:Sufficient}  is satisfied.
\end{proposition}
\textbf{Proof.}
As advertised, the approachability strategy we will consider is based on calibration (as it can be generalized to more complex settings). The main insight is that player~1 predicts, stage by stage, $y \in \Delta(\cJ)$ using a calibrated procedure and plays the response given by Lemma \ref{LM:eqBigMatch2}. Let $\theta_t$ be the sequence of weights used for the general payoff evaluation (recall that Cesaro average corresponds to $\theta_t=1/N$ while discounted evaluation is $\theta_t=\lambda(1-\lambda)^{t-1}$).

Let $\Big\{y_k, k \in \{1,\ldots,K\} \Big\}$ be a finite $\varepsilon$-discretization of $\Delta(\cJ)$ and $x_k$ be given by Lemma \ref{LM:eqBigMatch2} for every $ k \in \{1,\ldots,K\}$. Consider the calibration algorithm introduced in Lemma \ref{LM:Calib} with respect to the sequence of weights $\theta_t$. The strategy in the Big-Match game of type II is defined as follows: whenever $y_k \in \Delta(\cJ)$ is predicted by the calibration algorithm, player~1 actually plays accordingly to $x_k$. 

Assume that player~2 has never chosen an action in $\cJ^\star$ before stage $t$. Then Lemma \ref{LM:Calib} ensures that
\[\E\left[ \max_{k \in [K]}\Big(\sum_{s \in \N_t[k]} \theta_s\Big) \Big( \left\|y_k-\bar{\jmath}_t[k] \right\|-\varepsilon\Big)_+ \right] \leq \sqrt{8|\cJ|\sum_{s \leq t} \theta^2_s}
\]
where $\bar{\jmath}_t[k] \in \Delta(\cJ)$  is the weighted empirical distribution of actions of player~2 on the set of stages where $y_k$ was predicted. We recall that on each of these stages, player~1 played accordingly to $x_k\in\Delta(\cJ)$ so that the average weighted expected payoffs  on those stages is $g(x_k,\bar{\jmath}_t[k])$.

Summing over $k \in [K]$, we obtain that
\[d_\cC\left(\E\left[\frac{\sum_{k \in [K]}\Big(\sum_{s \in \N_t[k]} \theta_s\Big) g(x_k,\bar{\jmath}_t[k])}{\sum_{s\leq t} \theta_s}\right]\right) \leq \varepsilon+K\frac{\sqrt{8|\cJ|\sum_{s \leq t} \theta^2_s}}{\sum_{s\leq t} \theta_s}.
\]
We stress that the payoff on the left-hand side of the above equation is exactly the expected weighted average vectorial payoff obtained by player~1 up to stage $t$. %For instance, in the specific case where $\theta_s=\lambda(1-\lambda)^{s-1}$, the above equation  reads as\[ d_\cC\left(\E\left[\bar{g}_t^\lambda\right]\right) \leq \varepsilon + K\sqrt{8|\cJ|}\sqrt{\frac{\lambda}{2-\lambda}}\sqrt{\frac{1+(1-\lambda)^t}{1-(1-\lambda)^t}}.\]
As a consequence, if player~2 never uses quitting action in $\cJ^\star$, letting $t$ to infinity in the above equation yields \[d_\cC\left(\E\left[\bar{g}_\infty^\theta\right]\right) \leq \varepsilon+K\sqrt{8|\cJ|}\|\theta_2\|.\]
It remains to consider the case where player~2 used some quitting action $j^\star \in \cJ^\star$ at stage $\tau^\star+1 \in \N$. At that stage, player~1 played accordingly to  $x_k$ for some $k \in [K]$, which ensures that $g(x_k,j^\star) \in \cC$. As a consequence, absorption took place and the expected absorption payoff belongs to $\cC$. We therefore obtain that
 \begin{align*}d_\cC\left(\E\left[\bar{g}_\infty^\theta\right]\right) &\leq d_\cC\left( \frac{\sum_{s \leq \tau\star} \theta_s}{\sum_{s \in \N} \theta_s}  \bar{g}_{\tau^\star}^\theta+ (1-\frac{\sum_{s \leq \tau\star} \theta_s}{\sum_{s \in \N} \theta_s}  )g(x_k,j^\star) \right) \leq \frac{\sum_{s \leq \tau\star} \theta_s}{\sum_{s \in \N} \theta_s}d_\cC\left(\E\left[\bar{g}_{\tau^\star}^\theta\right]\right) \\
 &\leq \frac{\sum_{s \leq \tau^\star} \theta_s}{\sum_{s \in \N} \theta_s}\varepsilon+K\frac{\sqrt{8|\cJ|\sum_{s \leq \tau^\star} \theta^2_s}}{\sum_{s \in \N} \theta_s} \leq \varepsilon + K\sqrt{8|\cJ|}\|\theta\|_2,
\end{align*}
hence the result. \qed

\subsubsection{Condition \eqref{EQ:Sufficient} is sufficient in Big-Match games of type I} 
We now turn to the case of Big-match games of type I, where only player~1 can quit. In those games, we have the following useful equivalence result. %Condition \eqref{EQ:Sufficient} can also be rewritten in a more convenient, but less compact, way.
\begin{lemma}\label{LM:EqSufTypeI}
In Big-Match games of type I, Conditions \eqref{EQ:Sufficient}, \eqref{EQ:non-Necessary} and \eqref{EQ:Necessary} are all equivalent, and they are further equivalent to the usual Blackwell condition stated as
\begin{equation}
\label{B-cond} \forall \by \in \Delta(\bJ), \exists \bx \in \Delta(\bI), g(\bx,\by) \in \cC,
\end{equation} 
which also reads, equivalently, as
\begin{equation}\label{BMI-cond1}
\forall y \in \Delta(\cJ), \exists (x,x^\star,\gamma) \in \Delta(\cI) \times \Delta(\cI^\star) \times [0,1],  (1-\gamma) g(x,y) + \gamma g(x^\star,y) \in \cC.
\end{equation} 
\end{lemma}
A consequence of \eqref{BMI-cond1} is that if player~2 plays i.i.d.\ according to $y \in \Delta(\cJ)$, then player~1 can approach $\cC \subset \R^d$ by playing $x \in \Delta(\cI)$ ``perturbed''  by $x^\star \in \Delta(\cI^\star)$ with an overall total probability of absorption of $\gamma$.

\textbf{Proof:} We decompose the proof in three main parts. %First of all, in Big-Match games of type I, $\cJ^\star=\emptyset$ so $\bJ=\cJ$.

\textbf{Part a.} First we argue that Condition \eqref{EQ:Sufficient} implies the Blackwell condition \eqref{B-cond}. 
Decompose every $\bx \in \Delta(\bI)$ as $\bx=(1-\gamma_\bx) x + \gamma_\bx x^\star$, where $(x,x^\star,\gamma_\bx)\in \Delta(\cI) \times \Delta(\cI^\star) \times [0,1]$. Similarly, decompose every $\alpha \in \cM(\bI)$ into $\alpha_0 \in \cM(\cI)$ and $\alpha^\star \in \cM(\cI^\star)$. Then the fraction in Condition \eqref{EQ:Sufficient} can be rewritten into
\begin{equation}\label{newfraction} \frac{g(\bx,\by)+g(\alpha^\star,\by)+\gamma_\bx g(x^\star,\beta)}{1+\|\alpha^\star\|_1+\gamma_\bx \|\beta\|_1} \in \cC.
\end{equation}

Now suppose that $\cC$ satisfies Condition \eqref{EQ:Sufficient}. Let $\by\in\Delta(\bJ)$, and take an $\bx\in\Delta(\bI)$ that gives the minimum in Condition \eqref{EQ:Sufficient}. We distinguish two cases.

Suppose first that $\gamma_\bx >0$. Then, by taking $\beta_c = c\cdot \delta_{j}$ in \eqref{newfraction} and letting $c$ tend to infinity, we find $g(x^\star,j) \in \cC$ for all $j \in \cJ=\bJ$. Hence, $g(x^\star,\by)\in\cC$. %Consequently, player~1 can approach $\cC$ by playing $x^\star$, which induces absorption at the first stage. 

Now assume  that $\gamma_\bx=0$. Define $\hat{\bx}= \frac{\bx+\alpha^\star}{1+\|\alpha^\star\|_1} \in \Delta(\bI)$. Then in view of \eqref{newfraction}, we have 
\[g(\hat{\bx},\by)=\frac{g(\bx,\by)+g(\alpha^\star,\by)}{1+\|\alpha^\star\|_1 }\in\cC.
\]
So, in both cases, $\cC$ satisfies the Blackwell condition \eqref{B-cond}. 

\textbf{Part b.} Now we prove that the Blackwell condition \eqref{B-cond} implies Condition \eqref{EQ:Sufficient}. So, assume that $\cC$ satisfies \eqref{B-cond}. Then, for every $\by \in \Delta(\bJ)$, we decompose again the associated $\bx \in \Delta(\bI) $ as   $\bx=(1-\gamma_\bx) x + \gamma_\bx x^\star$. The choice of $(x,\alpha)$ where $\alpha = (\frac{1}{\gamma_\bx}-1)x^\star$ ensures that $p^\star(x,\beta)=0$ for all $\beta$, and hence $\cC$ satisfies Condition \eqref{EQ:Sufficient}.

\textbf{Part c.} We already know that Condition \eqref{EQ:Sufficient} implies Condition \eqref{EQ:non-Necessary}, which further implies Condition \eqref{EQ:Necessary}. Since Condition \eqref{EQ:Sufficient} is equivalent to the Blackwell condition \eqref{B-cond}, it only remains to verify that Condition \eqref{EQ:Necessary} implies \eqref{B-cond}. This can be easily checked by taking $\beta =0$ and using the same decomposition trick as above. \qed 

\begin{proposition}\label{PR:SuffType1} In Big-Match games of type I, a convex set $\cC$ is weakly approachable by player~1 if Condition \eqref{EQ:Sufficient}  is satisfied.
\end{proposition}
\textbf{Proof:} The approachability strategy is rather similar to the one introduced  for Big-Match games of type II. Given the finite $\varepsilon$-discretization of $\Delta(\cJ)$ denoted by $\{ y[k], k \in [K]\}$, Lemma \ref{LM:EqSufTypeI} guarantees, for any $\eta >0$, the existence of $x[k] \in \Delta(\cI)$, $x^\star[k]\in\Delta(\cI^\star)$ and $\gamma[k] \in [0,1-\eta]$ such that $(1-\gamma[k])g(x[k],y[k]) + \gamma[k]g(x^\star[k],y[k])$ is $2\eta$-close to $\cC$.

Based on an auxiliary calibration algorithm (to be adapted and described later) whose prediction at stage $\tau$ is some $y[k_\tau] \in \Delta(\cJ)$, we consider the strategy of player~1 that dictates to play at this stage
\begin{align*} 
& x^\star[k_\tau] \ \text{ with probability }\ \gamma_\tau[k_\tau]:=\frac{\gamma[k_\tau] \theta_\tau}{(1-\gamma[k_\tau])\sum_{s=\tau}^\infty \theta_s+\gamma[k_\tau] \theta_\tau}\\
& \text{and } x[k_\tau] \ \text{ with probability }\ 1-\gamma_\tau[k_\tau] \ .
\end{align*}
Thus at stage $\tau$, with probability $\gamma_\tau[k_\tau]$ player~1 quits according to $x^\star[k_\tau]$ (then the expected absorption payoff is $g(x^\star[k_\tau],j_\tau)$), and with the remaining probability, which is positive, he plays $x[k_\tau]$ and play does not absorb at this stage. Since the cumulative weight of all the remaining stages is $\sum_{s=\tau}^\infty \theta_s$, the associated expected payoff of the decision taken at stage $\tau$ is
\begin{align*} &\gamma_\tau[k_\tau]\sum_{s=\tau}^\infty \theta_s  g(x^\star[k_\tau],j_\tau)  + (1-\gamma_\tau[k_\tau]) \theta_\tau g(x[k_\tau],j_\tau) \\ &= \frac{\theta_\tau \sum_{s=\tau}^\infty \theta_s}{(1-\gamma[k_\tau])\sum_{s=\tau}^\infty \theta_s+\gamma[k_\tau] \theta_\tau} \bigg(\gamma[k_\tau]g(x^\star[k_\tau],j_\tau)+(1-\gamma[k_\tau])g(x[k_\tau],j_\tau)\bigg)\\
&= \theta^\star_\tau \Big(\gamma[k_\tau]g(x^\star[k_\tau],j_\tau)+(1-\gamma[k_\tau])g(x[k_\tau],j_\tau)\Big),
\end{align*}
where $\theta^\star_\tau:=\theta_\tau (1-\gamma_\tau[k_\tau])/(1-\gamma[k_\tau])$. As a consequence, summing over $\tau \in \N$ and using the fact that the game is absorbed at stage $\tau$ with probability $\gamma_\tau[k_\tau]$, we obtain that
\begin{align*} \E\left[\overline{g}^\theta_\infty\right]&=\sum_{\tau=1}^\infty  \prod_{s <\tau} \big( 1-\gamma_s[k_s]\big)\theta^\star_\tau \Big(\gamma[k_\tau]g(x^\star[k_\tau],j_\tau)+(1-\gamma[k_\tau])g(x[k_\tau],j_\tau)\Big) \\
& := \sum_{\tau=1}^\infty \hat{\theta}_\tau \Big(\gamma[k_\tau]g(x^\star[k_\tau],j_\tau)+(1-\gamma[k_\tau])g(x[k_\tau],j_\tau)\Big) .
\end{align*}
We stress the fact that the sequence $\hat{\theta}_\tau$ is predicable with respect to the filtration induced by the strategies of players 1 and 2 (i.e., it does not depend  on the choices made at stages $s \geq \tau +1$).
\medskip

To define the strategy of player~1, we consider an auxiliary algorithm calibrated with respect to the sequence of weights $\hat{\theta}_\tau$ (which is possible even though $\hat{\theta}_\tau$ depends on the past predictions).
Using the fact that $\hat{\theta}_s\leq \theta^\star_s \leq \theta_s/\eta$,  Lemma \ref{LM:Calib} and the same argument as in Big-match games of type II, we obtain that our strategy guarantees the following:
\[d_\cC\left(\E\left[\overline{g}^\theta_\infty\right]\right) \leq 2\eta+\frac{\varepsilon}{\eta} + K \frac{\sqrt{8|\Omega|}}{\eta}\|\theta\|_2.\]
The result follows by, for instance, taking $\eta=\sqrt{\varepsilon}$.\qed

\subsubsection{Condition \eqref{EQ:Sufficient} is sufficient in all generalized quitting games}
Using the tools introduced in the previous subsections for Big-Match games of types I and II, we are now able to give a simple proof of the main result, that Condition \eqref{EQ:Sufficient} is sufficient to ensure weak approachability in all generalized quitting games.

We start with a useful consequence of Condition \eqref{EQ:Sufficient}.
\begin{lemma}\label{LM:EqSufQuitting}
In generalized quitting games, Condition \eqref{EQ:Sufficient} implies that at least one of the following conditions holds:
\begin{enumerate}
\item[(a)] $\exists(x,x^\star,\gamma) \in \Delta(\cI) \times \Delta(\cI^\star) \times (0,1]$ with
\begin{equation} \label{alt1}
g(x^\star,j) \in \cC, \ \forall j \in \cJ\quad \text{ and }\quad
\ g( (1-\gamma)x + \gamma x^\star,j^\star) \in \cC,  \forall j^\star \in \cJ^\star.
\end{equation}
\item[(b)] $\forall y \in \Delta(\cJ), \exists (x,x^\star,\gamma) \in \Delta(\cI) \times \Delta(\cI^\star) \times [0,1]$ with   
\begin{equation} \label{alt2}
\hspace{-1.6cm}g(x,j^\star) \in \cC,  \forall j^\star \in \cJ^\star \quad \text{ and }\quad \ (1-\gamma) g(x,y) + \gamma g(x^\star,y) \in \cC.
\end{equation}
\end{enumerate}
\end{lemma}
\textbf{Proof:} Let $y\in\Delta(\bJ)$. Assume first that Condition \eqref{EQ:Sufficient} is satisfied with some $\bx \in \Delta(\bI)$ that puts positive weight on $\cI^\star$, i.e. $\bx=(1-\gamma )x+\gamma x^\star$ with $\gamma \in (0,1]$. Then, taking $\beta=c\cdot\delta_j$ in the expression of Condition \eqref{EQ:Sufficient} with $c$ going to infinity, for any $j\in \bJ$, yields condition (a).

Otherwise, Condition \eqref{EQ:Sufficient} is satisfied for some $x \in \Delta(\cI)$. Then, the same argument as in the proof of Lemma \ref{LM:EqSufTypeI} yields condition (b).
\qed
\begin{proposition}\label{proplabel} In generalized quitting games, a convex set $\cC$ is weakly approachable by player~1 if Condition \eqref{EQ:Sufficient}  is satisfied.
\end{proposition}
\textbf{Proof.} Assume that Condition \eqref{EQ:Sufficient} is satisfied. Then either condition (a) or condition (b) of Lemma \ref{LM:EqSufQuitting} is satisfied.

First assume that condition (a) of Lemma \ref{LM:EqSufQuitting} is satisfied. Then, player~1 just has to play i.i.d.\ according to $(1-\gamma)x+\gamma x^\star \in \Delta(\bI)$. Indeed, then the probability of absorption at each stage is at least $\gamma$, so absorption will eventually take place with probability 1, and by condition (a) the expected absorption payoff is in $\cC$. As a consequence,
\[d_\cC\left(\E\left[\overline{g}^\theta_\infty\right]\right) \leq \sum_{s=1}^\infty (1-\gamma)^s \theta_s \leq \frac{1-\gamma}{\sqrt{2\gamma-\gamma^2}} \|\theta\|_2,\]
hence the result.

Now assume that condition (b) of Lemma \ref{LM:EqSufQuitting} is satisfied. We claim that the strategy defined in the proof of Proposition \ref{PR:SuffType1} is an approachability strategy. Indeed, as long as player~2 does not play an absorbing action $j^\star\in \cJ^\star$, the analysis is identical.

If, on the other hand, player~2 plays $j^\star \in \cJ^\star$ at stage $\tau^\star$, then the absorbing payoff is equal to
\[
g((1-\gamma_{\tau^\star}[k_{\tau^\star}])x[k_{\tau^\star}] +\gamma_{\tau^\star}[k_{\tau^\star}]x^\star[k_{\tau^\star}],j^\star)\]
\[ = g(x[k_{\tau^\star}],j^\star)+ \gamma_{\tau^\star}[k_{\tau^\star}]\Big( g(x^\star[k_{\tau^\star}],j^\star)-g(x[k_{\tau^\star}],j^\star)\Big),\]
which is therefore within a distance of $2\gamma_{\tau^\star}[k_{\tau^\star}]$ to $\cC$. As a consequence,
\begin{align*}d_\cC\left(\E\left[\overline{g}^\theta_\infty\right]\right) &\leq \E\left[\eta+\frac{\varepsilon}{\eta} + K \sqrt{8|\Omega|\sum_{s < {\tau^\star}} \hat{\theta}_s^2} + 2 \gamma_\tau[k_{\tau^\star}] \sum_{s={\tau^\star}}^\infty \theta_s\right] \\ &\leq \eta+\frac{\varepsilon}{\eta} + K \frac{\sqrt{8|\Omega|\sum_{s \in \N} \theta_s^2}}{\eta} + 2 \E\left[\frac{\theta_{\tau^\star}}{\eta}\right] \\&\leq\eta+\frac{\varepsilon}{\eta} + K \frac{\sqrt{8|\Omega|}}{\eta}\|\theta\|_2 + 2 \frac{\|\theta\|_2}{\eta}.
\end{align*}
And the result follows, by taking $\eta=\sqrt{\varepsilon}$. \qed

\section{Weak vs Uniform Approachability}

In this section, we compare the notions of weak and uniform approachability in Big-Match games. 
%of type II. Then, we prove that in generalized quitting games, the sufficient Condition \eqref{EQ:Sufficient} is not necessary for weak approachability and, reciprocally, that Condition \eqref{EQ:Necessary} is not sufficient. The fact that Condition \eqref{EQ:non-Necessary} is neither sufficient nor necessary for weak approachability (even in Big-match games of type II) is delayed to the Appendix (the proof obviously relies on counter-examples).

First we consider Big-Match games of type I. In this class of games, there are convex sets that are weakly approachable but not uniformly approachable, which is illustrated by Example \ref{ex4} in Section \ref{Sec-Ex}. Hence, in this class of games, uniform determinacy fails. In contrast, weak determinacy holds true by Proposition \ref{MProp2}. 

Now we turn our attention to Big-Match games of type II. We have the following characterization result for uniform approachability.

\begin{proposition}\label{Prop:UnifType2}
In Big-Match games of type II, a convex set $\cC$ is uniformly approachable by player~1 if and only if Condition \eqref{EQ:Sufficient} holds.
%However, there are convex sets that are neither anytime approachable nor anytime excludable.
\end{proposition}
In view of Lemma \ref{LM:eqBigMatch2}, for Big-Match games of type II, Condition \eqref{EQ:Sufficient} can be rewritten as
\begin{equation}
\label{withoutabs}
\forall \by \in \Delta(\bJ), \exists x \in \cX_\cC, g(x,\by) \in \cC,
\end{equation}
where 
$$\mathcal{X}_\cC =\{x \in \Delta(\cI) \ \text{ s.t. } \ g(x,j^\star)\in \cC, \forall j^\star \in \cJ^\star\}.$$
Note that \eqref{withoutabs} is exactly Blackwell's approachability condition for the set $\cC$ in the following related repeated game (with no absorption): player~1 is restricted to play mixed actions in the set $\mathcal{X}_\cC$, player~2 can choose actions in $\bJ$, and the payoff is given by $g$.

\textbf{Proof:} The sufficiency part is a direct consequence of Proposition \ref{lem-BMI-suff}.
%First, assume that $\cC$ satisfies \eqref{withoutabs}. Then, player~1 can uniformly approach $\cC$ by a strategy $\widetilde{\sigma}$ in the related repeated game (without absorption). This strategy naturally induces a strategy $\sigma$ in the Big-Match game of type II. Suppose that player~1 uses it. If player~2 chooses an action in $\cJ^\star$ at some stage, then it follows from the definition of $\cX_\cC$ that the expected absorption payoff is in $\cC$ (and thus the limit payoff). Otherwise, if player~2 only uses actions in $\cJ$, then the play never absorbs and the choice of $\sigma$ guarantees that, asymptotically, the expected average payoff is in $\cC$. Hence, player~1 can uniformly approach $\cC$ by $\sigma$.

Now assume that  $\cC$ does not satisfy \eqref{withoutabs}, i.e., there exists $\by_0\in\Delta(\bJ)$ such that $g(x,\by_0) \not\in \cC$ for all $x \in \cX_\cC$. By the definition of $\cX_\cC$, this implies that there exists $y_0 \in \Delta(\cJ)$ such that $g(x,y_0) \not\in\cC$ for all $x \in \cX_\cC$. Since $\cX_\cC$ is compact, there exists $\delta>0$ such that $d_\cC( g(x,y_0)) \geq \delta$ for all $x \in \cX_\cC$. By continuity, this also implies that  $d_\cC( g(x,y_0)) \geq \delta/2$ for all mixed actions $x \in \cX_\cC^\eta$, the $\eta$-neighborhood of $\cX_\cC$, for $\eta>0$ small enough. By continuity, there also exists $\varepsilon>0$ such that for all $x \not\in \cX_\cC^\eta$, there exists $j^\star \in \cJ^\star$ with $d_\cC(g(x,j^\star)) \geq \varepsilon$.

In conclusion, if player~1 only uses mixed actions in $\cX_\cC^\eta$ then $y_0$ ensures  that the average payoff is asymptotically $\delta/2$-away from $\cC$, whereas if player~1 uses, at some stage $t$,  a mixed action $x_t \not\in \cX_\cC^\eta$ then there is $j^\star \in \cJ^\star$ such that $g(x_t,j^\star)$ is $\varepsilon$-away from $\cC$. Thus, $\cC$ is not uniformly approachable.\qed

\medskip

In Big-Match games of type II, uniform determinacy fails: there are convex sets that are neither uniformly approachable nor uniformly excludable. This is demonstrated by Example \ref{ex5} in Section \ref{Sec-Ex}. The question remains open whether weak determinacy holds true for Big-Match games of type II.

\section{Examples}\label{Sec-Ex}
In this section we consider a number of examples. In all these examples, player~1 is the row player and player~2 is the column player. Quitting actions of the players are marked with a superscript $\star$. The payoffs are given by the corresponding matrices.

\begin{example}\label{ex1}\rm

The following generalized quitting game shows that Condition \eqref{EQ:Sufficient} is generally not necessary for weak (and hence for uniform) approachability.

\[\begin{tabular}{l|l|c|}
\multicolumn{1}{r}{}
 &  \multicolumn{1}{c}{$L^\star$}
 & \multicolumn{1}{c}{$R$} \\
\cline{2-3}
$T^\star$ & $1$ & $1$ \\
\cline{2-3}
$B$ & $-1$ & $-1$ \\
\cline{2-3}
\end{tabular}\vspace{0.1cm} \]
In this game, actions $T$ and $L$ are quitting, whereas actions $B$ and $R$ are non-quitting. 

The set $\cC=\{0\}$ does not satisfy Condition \eqref{EQ:Sufficient}. Indeed, the distance in \eqref{EQ:Sufficient} can be made arbitrary close to 1 by player~2 as follows. Choose any $\by\in\Delta(\bJ)$. Then given $\bx\in\Delta(\bI)$ and $\alpha\in \cM(\bI)$, choose $\beta=(0,r)$ with a large $r$ if $\bx$ puts a positive weight on action $T$, and choose $\beta=(r,0)$ with a large $r$ otherwise. This choice of $\beta$ implies that the fraction in \eqref{EQ:Sufficient} is close to 1 in the former case (due to the absorption payoff of 1 in entry $(T,R)$), and close to $-1$ in the latter case (due to the absorption payoff of $-1$ in entry $(B,L)$). In either case, the distance in \eqref{EQ:Sufficient} is close to 1.

Yet, playing $1/2T+1/2B$ at the first stage and $B$ at all remaining stages approaches  $\{0\}$, both weakly and uniformly. 

We remark that Condition \eqref{EQ:Sufficient} is generally not necessary for weak (and hence uniform) approachability even in Big-Match games of type II. This is shown later by Example \ref{ex5}, which involves a more difficult proof. $\Diamond$
\end{example}

\begin{example}\label{ex2}\rm
The next game demonstrates that weak (and hence uniform) determinacy fails in generalized quitting games, and that Conditions \eqref{EQ:non-Necessary} and \eqref{EQ:Necessary} are generally not sufficient for weak (and hence uniform) approachability.

\[\begin{tabular}{l|l|c|}
\multicolumn{1}{r}{}
 &  \multicolumn{1}{c}{$L^\star$}
 & \multicolumn{1}{c}{$R^\star$} \\
\cline{2-3}
$T^\star$ & $1$ & $0$ \\
\cline{2-3}
$B^\star$ & $0$ & $-1$ \\
\cline{2-3}
\end{tabular}\vspace{0.1cm} \]
In this game all actions are quitting. Note that this game is not a Big-Match game, as neither player has a non-quitting action. (Later, we will see versions of this game when some of the actions are non-quitting instead -- see Examples \ref{ex4} and \ref{ex6}.)

The set $\cC=\{0\}$ satisfies Conditions \eqref{EQ:non-Necessary} and \eqref{EQ:Necessary}, due to the fact that $\alpha$ is the last in the order of terms in \eqref{EQ:non-Necessary} and \eqref{EQ:Necessary}. Yet, $\{0\}$ is trivially not approachable and not excludable, neither weakly nor uniformly. 

Note that Blackwell's approachability condition is actually satisfied in this game: against a strategy that plays $y L + (1-y) R$ at the first stage, player~1 can ensure by playing $(1-y)T+yB$ that the expected payoff is at all stages equal to 0. 

We remark that Condition \eqref{EQ:Necessary} is generally not sufficient for weak (and hence uniform) approachability even in Big-Match games of type II. This is shown later by Example \ref{ex6}, which involves a more difficult proof. $\Diamond$
\end{example}

%\begin{example}\label{ex3}\rm
%The game below arises by adding a non-quitting action to Example \ref{ex2}.  
%
%\[\begin{tabular}{l|c|c|}
%\multicolumn{1}{r}{}
 %&  \multicolumn{1}{c}{$L^\star$}
 %& \multicolumn{1}{c}{$R^\star$} \\
%\cline{2-3}
%$T^\star$ & $1$ & $0$ \\
%\cline{2-3}
%$B^\star$ & $0$ & $-1$ \\
%\cline{2-3}
%$D$ & $-1$ & $-1$ \\
%\cline{2-3}
%\end{tabular}\vspace{0.1cm} \]
%This game is a Big-Match game of type I. 
%
%The new action $D$ increases player~1's possibilities tremendously, as it gives him the option to wait and see how player~2 behaves. Indeed, consider again the set $\cC=\{0\}$. In contrast with Example \ref{ex2}, in this game the set $\{0\}$ is approachable by player~1. This follows from the fact that $\{0\}$ satisfies Condition \eqref{EQ:Sufficient}, and therefore Proposition \ref{PR:SuffType1} applies. $\Diamond$
%\end{example}

\begin{example}\label{ex4}\rm
The following game shows that, in Big-Match games of type I, there are convex sets that are weakly approachable but not uniformly approachable, and thus uniform determinacy fails (a property already noted by S.\ Sorin).

\[\begin{tabular}{l|l|c|}
\multicolumn{1}{r}{}
 &  \multicolumn{1}{c}{$L$}
 & \multicolumn{1}{c}{$R$} \\
\cline{2-3}
$T^\ast$ & $1$ & $0$ \\
\cline{2-3}
$B$ & $0$ & $-1$ \\
\cline{2-3}
\end{tabular}\]\vspace{0.15cm}

Note that the payoffs are identical to those in Example \ref{ex2}. However, action $B$ in this game is non-quitting, which increases player~1's possibilities tremendously. Indeed, playing action $B$ gives player~1 the option to wait and see how player~2 behaves.

Consider the set $\cC=\{0\}$. This set satisfies Condition \eqref{EQ:Sufficient}, which follows easily from Lemma \ref{LM:EqSufTypeI}. Hence, by Proposition \ref{PR:SuffType1}, $\{0\}$ is weakly approachable by player~1. This also implies that $\{0\}$ is not uniformly excludable by player~2.

The set $\{0\}$ is however not uniformly approachable. The main reason is that player~1 cannot use the quitting action $T$ effectively, since it backfires if player~2 places positive probability on action $L$. A precise argument is given later in Proposition \ref{prop-bmI}. Hence, the game is not uniformly determined. $\Diamond$
\end{example}

\begin{example}\label{ex5}\rm
The following game shows that, in Big-Match games of type II, Conditions \eqref{EQ:Sufficient} and \eqref{EQ:non-Necessary} are not necessary for weak approachability, and that there convex sets that are weakly approachable but not uniformly approachable, and thus uniform determinacy fails.
\[\begin{tabular}{l|l|c|}
\multicolumn{1}{r}{}
 &  \multicolumn{1}{c}{$L^\star$}
 & \multicolumn{1}{c}{$R$} \\
\cline{2-3}
$T$ & $1$ & $1$ \\
\cline{2-3}
$B$ & $0$ & $-1$ \\
\cline{2-3}
\end{tabular}\]\vspace{0.1cm}

Consider the set $\cC=\{0\}$. Conditions \eqref{EQ:Sufficient} and \eqref{EQ:non-Necessary} are not satisfied for $\{0\}$, which can be easily verified with the help of Lemma \ref{LM:eqBigMatch2}. Indeed, for $y=(1/2,1/2)$, there is no $x\in\Delta(\cI)$ satisfying \eqref{suff-BMII}. Yet, as we show in Appendix A.2, player~1 can weakly approach $\cC$. This shows that Conditions \eqref{EQ:Sufficient} and \eqref{EQ:non-Necessary} are not necessary for weak approachability in Big-Match games of type II. 

In view of Proposition \ref{Prop:UnifType2}, the set $\{0\}$ is not uniformly approachable by player~1. Since $\{0\}$ is weakly approachable by player~1, we also obtain that $\{0\}$ is not uniformly excludable by player~2. Hence, the game is not uniformly determined. $\Diamond$
\end{example}

\begin{example}\label{ex6}\rm
The following game shows that, in Big-Match games of type II, Condition \eqref{EQ:Necessary} is not sufficient for weak approachability. 

\[\begin{tabular}{l|l|c|}
\multicolumn{1}{r}{}
 &  \multicolumn{1}{c}{$L^\star$}
 & \multicolumn{1}{c}{$R$} \\
\cline{2-3}
$T$ & $1$ & $0$ \\
\cline{2-3}
$B$ & $0$ & $-1$ \\
\cline{2-3}
\end{tabular}\]\vspace{0.2cm}

Consider the set $\cC=\{0\}$. We argue that Condition \eqref{EQ:Necessary} is satisfied. Indeed, take any $\by\in \Delta(\bJ)$ and $\beta\in\cM(\bJ)$. Then by choosing $$\bx=\left(\frac{\by_R}{1+\beta_L},\frac{\by_L+\beta_L}{1+\beta_L}\right)\hspace{0.4cm}\text{and}\hspace{0.4cm}\alpha=(0,0),$$
we obtain
$$g(\bx,\by)+g^*(\alpha,\by)+g^*(\bx,\beta)=\frac{\by_R}{1+\beta_L}\cdot (\by_L+\beta_L)-\frac{\by_L+\beta_L}{1+\beta_L}\cdot \by_R=0,$$
which implies that Condition \eqref{EQ:Necessary} holds indeed.

Clearly, Conditions \eqref{EQ:Sufficient} and \eqref{EQ:non-Necessary} are not satisfied, which can be easily verified with the help of Lemma \ref{LM:eqBigMatch2}. We will show in Appendix A.1 that this prevents the weak approachability of $\{0\}$. $\Diamond$
\end{example}

%To understand the difficulty, we refer the reader to the appendix.

\section{Conclusion}
We have introduced the model of stochastic games with vector payoffs and have exhibited a sufficient condition and a strongly related necessary condition for the weak approachability of a convex set in the class of generalized quitting games. In Big-Match games of type I the conditions coincide, but generally they differ in Big-Match games of type II. 

Some of our conditions are also useful for uniform approachability, though in a non-obvious way. In fact, as we have seen, weak and uniform approachability conditions drastically differ, even in Big-Match games. 

When Condition \eqref{EQ:Sufficient} is satisfied,  we have also provided explicit strategies for weak approachability based on an auxiliary calibration strategy (itself induced by a traditional approachability result). The question of optimal rates of convergence and efficient algorithms are left open:  our techniques provide qualitative results, and, unfortunately the rates  decrease with the dimension, as we need to consider $\varepsilon$-discretization of $\Delta(\cJ)$.

\bibliographystyle{plain}
\bibliography{ApproachAbsorbing}

 \appendix 
 
\section{On weak approachability in Big-Match games of type II}
We have proved that Condition \eqref{EQ:Sufficient} and Condition \eqref{EQ:Necessary} are respectively sufficient and necessary for the weak approachability of a convex target set $\cC \subset \R^d$ in generalized quitting games (cf. Theorem \ref{TH: Results}). Even though these two conditions are equivalent in Big-Match games of type I (cf. Lemma \ref{LM:EqSufTypeI}), generally there is still a gap between them. 

The goal of this section is to provide further insight by considering specific examples in the calls of Big-Match games of type II. Recall that, in this class, Conditions \eqref{EQ:Sufficient} and \eqref{EQ:non-Necessary} are equivalent (cf. Lemma \ref{LM:eqBigMatch2}).

In the next subsections, we prove that Condition \eqref{EQ:Necessary} is not sufficient and that Conditions \eqref{EQ:Sufficient} and \eqref{EQ:non-Necessary} are not necessary for weak approachability in Big-Match games of type II. We then provide some necessary and sufficient conditions, in the very specific case where player~2 has only one absorbing and one non-absorbing action, based on techniques in continuous time, similar to the one used by Vieille~\cite{Vie92} when he proves weak determinacy for non convex sets in the classical Blackwell setting.

\subsection{Condition \eqref{EQ:Necessary} is not sufficient for weak approachability}

We revisit the Big-Match game of type II from Example \ref{ex6}.

\[\begin{tabular}{l|l|c|}
\multicolumn{1}{r}{}
 &  \multicolumn{1}{c}{$L^\star$}
 & \multicolumn{1}{c}{$R$} \\
\cline{2-3}
$T$ & $1$ & $0$ \\
\cline{2-3}
$B$ & $0$ & $-1$ \\
\cline{2-3}
\end{tabular}\]\vspace{0.2cm}

Consider again the set $\cC=\{0\}$. We will now show that $\{0\}$ is not weakly approachable. 

For the sake of simplicity, we will consider $\lambda$-discounted payoffs. Given the strategies $\sigma$ and $\tau$, we denote 
$$\bar{g}^\lambda_\infty(\sigma,\tau)=\mathbb{E}_{(\sigma,\tau)}\left(\sum_{t=1}^\infty \la (1-\la)^{j-1} g(i_t,j_t)\right) = \sum_{t=1}^\infty \la (1-\la)^{t-1} \mathbb{E}_{\sigma,\tau} g(i_t,j_t).$$
Similarly, the expected $\la$-discounted payoff up to period $k$ is
$$\bar{g}^\lambda_k(\sigma,\tau)=\mathbb{E}_{(\sigma,\tau)}\left(\sum_{t=1}^k \la (1-\la)^{t-1} g(i_t,j_t)\right) = \sum_{t=1}^k \la (1-\la)^{t-1} \mathbb{E}_{\sigma,\tau} g(i_t,j_t).$$

\begin{proposition}
There exists a $\lambda^\star\in(0,1)$ such that, for every $\lambda\in(0,\lambda^\star)$ and every strategy $\sigma$ of player~1, there is a strategy $\tau$ of player~2 with the property that $\bar{g}^\lambda_\infty(\sigma,\tau)\not\in [-\tfrac{1}{2e},\tfrac{1}{2e}]$.
\end{proposition}

\noindent Consequently, Condition \eqref{EQ:Necessary} is not sufficient for weak approachability.

\textbf{Idea of the proof: } Denote $\ep=\tfrac{1}{2e}$, and consider a small discount factor $\la\in(0,1)$. The main problem for player~1 is the following. Player~1 needs to guarantee an expected $\la$-discounted payoff close to 0 against the strategy $R^\infty$ of player~2, which means that he has to put large probabilities on action $T$ at some point. However, a large probability on $T$ could easily backfire if player~2 plays $L$ instead. As our analysis will show, player~1's best chance is to gradually increase the probability on $T$. The maximal probability that player~1 can put on $T$ at period 1 is $\ep$, since higher probabilities would lead to an expected $\la$-discounted payoff higher than $\ep$ if player~2 plays $L$. Thus, player~1 should play $T$ with probability $\ep$ at period 1, and this is safe against action $L$. If player~2 plays $R$ at period 1, then the expected payoff at period 1 is negative, $-1+\ep$ to be precise, and this allows player~1 to increase the probability on action $T$ up to $\ep+\tfrac{\la}{1-\la}$ at period 2. Indeed, if player~2 plays $R$ at period 1 and plays $L$ at period 2, then the expected $\la$-discounted payoff is exactly $\ep$. By continuing so, we obtain a sequence $(\oz_{\la k})_{k\in\dN}$ for the probabilities on $T$. This sequence is strictly increasing until it reaches 1 and then it stays 1 forever. Let $\sigma^\star_\la$ denote the corresponding Markov strategy for player~1 which uses these probabilities on $T$ during the game. We will show that this strategy $\sigma^\star_\la$ is indeed player~1's best chance. However, for small $\la\in(0,1)$, the probabilities on $T$ do not converge fast enough to 1 and consequently, when player~2 always plays action $R$, the expected $\la$-discounted payoff stays below $-\ep$.\medskip

\noindent\textbf{Step 1: A reduction.} We argue that it is sufficient to prove the statement of the proposition when we only consider Markov strategies (i.e. strategies where the probabilities on the actions depend only on the stage). That is, if there exists a $\lambda_M\in(0,1)$ such that, for every $\lambda\in(0,\lambda_M)$ and every Markov strategy $\sigma$ for player~1, there is a Markov strategy $\tau$ for player~2 with the property that $\bar{g}^\lambda_\infty(\sigma,\tau)\not\in [-\ep,\ep]$, then the proposition follows with $\lambda^\star=\la_M$.\medskip

\noindent\textbf{Proof for step 1: } Assume that such a $\la_M\in(0,1)$ exists. Consider an arbitrary strategy $\sigma'$ for player~1. For every $k\in\mathbb{N}$, let $p_k$ denote the probability, with respect $\sigma'$ and $R^\infty$, that player~1 plays action $T$ at period $k$. Now define the Markov strategy $\sigma$ for player~1 which prescribes to play action $T$ with probability $p_k$ at every period $k\in\mathbb{N}$. Let $\la\in(0,\la_M)$. Since $\sigma$ is a Markov strategy, by our assumption there exists a Markov strategy $\tau$ for player~2 with the property that $\bar{g}^\lambda_\infty(\sigma,\tau)\not\in [-\ep,\ep]$. Because $(\sigma',\tau)$ and $(\sigma,\tau)$ generate the same expected payoff for each period $k\in\mathbb{N}$, we have $\bar{g}^\lambda_\infty(\sigma',\tau)=\bar{g}^\lambda_\infty(\sigma,\tau)$. Hence, $\bar{g}^\lambda_\infty(\sigma',\tau)\not\in [-\ep,\ep]$, and the proposition will then follow by choosing $\lambda^\star=\la_M$.\medskip

\noindent\textbf{Step 2: The main strategy $\sigma^\star_\la$.} Define 
$$z_{\la k}=\ep+(k-1)\frac{\la}{1-\la}\hspace{0.5cm}\text{and}\hspace{0.5cm}\oz_{\la k}=\min\{z_{\la k},1\}$$
for every $\la\in(0,1)$ and every $k\in\dN$. For every $\la\in(0,1)$, the sequence $(z_{\la k})_{k=1}^\infty$ is positive and strictly increasing, and it diverges to infinity. So, for every $\la\in(0,1)$, there is a unique $k_\la\in\dN$ such that $z_{\la k_\la}\leq 1 < z_{\la, k_\la+1}$. 

For every $\la\in(0,1)$, let $\sigma^\star_\la$ be the Markov strategy for player~1 which prescribes to play action $T$ with probability $\oz_{\la k}$ at every period $k\in\dN$. We argue that for every $\la\in(0,1)$
\begin{equation}\label{eqR}
\bar{g}^\lambda_\infty(\sigma^\star_\la,R^\infty)=\ep - \la(1-\la)^{k_\la-1}-(1-\la)^{k_\la} z_{\la ,k_\la}<\ep-(1-\la)^{k_\la},
\end{equation}
that for every $\la\in(0,1)$ and $k\in\{1,\ldots,k_\la\}$
\begin{equation}\label{eqL[k].1}
\bar{g}^\lambda_\infty(\sigma^\star_\la,L[k])=\ep,
\end{equation}
and that for every $\la\in(0,1)$ and $k\in\dN$ with $k>k_\la$
\begin{equation}\label{eqL[k].2}
\bar{g}^\lambda_\infty(\sigma^\star_\la,L[k])<\ep.
\end{equation}\medskip

\noindent\textbf{Proof for step 2: } Fix an arbitrary $\la\in(0,1)$. We first prove 
\begin{equation}\label{iter}
\bar{g}^\lambda_{k}(\sigma^\star_\la,R^\infty)=\ep - \la(1-\la)^{k-1}-(1-\la)^k z_{\la k}
\end{equation}
for every $k\in\{1,\ldots,k_\la\}$ by induction on $k$. For $k=1$ we have 
$$\bar{g}^\lambda_1(\sigma^\star_\la,R^\infty)=\la (0\cdot z_{\la 1}-1\cdot (1-z_{\la 1}))=\la(-1+\ep)$$
and
$$\ep - \la-(1-\la) z_{\la 1}=\ep-\la-(1-\la)\ep=\la(-1+\ep),$$
hence (\ref{iter}) is true for $k=1$. Now, suppose that (\ref{iter}) is true for some $k$ and that $k+1\leq k_\la$. Then
\begin{eqnarray*}
\bar{g}^\lambda_{k+1}(\sigma^\star_\la,R^\infty) &=& \bar{g}^\lambda_{k}(\sigma^\star_\la,R^\infty) + \la(1-\la)^k(-1+z_{\la ,k+1}) \\
 &=& \ep - \la(1-\la)^{k-1}-(1-\la)^k z_{\la k} + \la(1-\la)^k(-1+z_{\la ,k+1})\\
 &=& \ep(1-(1-\la)^{k+1}) - \la(1-\la)^{k-1}-(k-1)\la(1-\la)^{k-1}-\la(1-\la)^k+k\la^2(1-\la)^{k-1}\\
 &=& \ep(1-(1-\la)^{k+1})-\la(1-\la)^k-k\la(1-\la)^{k-1}+k\la^2(1-\la)^{k-1}\\
 &=& \ep(1-(1-\la)^{k+1})-\la(1-\la)^k-k\la(1-\la)^k\\
 &=& \ep - \la(1-\la)^k-(1-\la)^{k+1} z_{\la ,k+1},
\end{eqnarray*}
which proves (\ref{iter}) for $k+1$. So we have shown (\ref{iter}) for every $k\in\{1,\ldots,k_\la\}$.

Since $\sigma^\star_\la$ prescribes action $T$ with probability 1 from period $k_\la+1$ onwards, (\ref{iter}) for $k_\la$ implies 
\begin{eqnarray*}
\bar{g}^\lambda_\infty(\sigma^\star_\la,R^\infty) &=& \bar{g}^\lambda_{k_\la}(\sigma^\star_\la,R^\infty)\\
 &=& \ep - \la(1-\la)^{k_\la-1}-(1-\la)^{k_\la} z_{\la ,k_\la}\\
 &<& \ep - \la(1-\la)^{k_\la-1}-(1-\la)^{k_\la} (1-\tfrac{\la}{1-\la})\\
 &=& \ep-(1-\la)^{k_\la},
\end{eqnarray*}
which proves (\ref{eqR}).

In view of (\ref{iter}), for any $k\in\{1,\ldots,k_\la\}$ we have
\begin{eqnarray}
\bar{g}^\lambda_{k}(\sigma^\star_\la,R^\infty)+(1-\la)^k z_{\la ,k+1} &=& \ep - \la(1-\la)^{k-1}-(1-\la)^k z_{\la k} + (1-\la)^k z_{\la ,k+1}\nonumber\\
 &=& \ep - \la(1-\la)^{k-1} - (k-1)\la(1-\la)^{k-1} + k\la(1-\la)^{k-1}\nonumber\\
 &=& \ep.\label{=ep}
\end{eqnarray}
Note that $\bar{g}^\lambda_\infty(\sigma^\star_\la,L[1])=z_{\la 1}=\ep$, and for every $k\in\{2,\ldots,k_\la\}$ we have by (\ref{=ep}) that 
$$\bar{g}^\lambda_\infty(\sigma^\star_\la,L[k]) = \bar{g}^\lambda_{k-1}(\sigma^\star_\la,R^\infty)+(1-\la)^{k-1} z_{\la k} =\ep.$$
Hence, we have proven (\ref{eqL[k].1}).

Finally, assume that $k\in\dN$ with $k>k_\la$. Since $\sigma^\star_\la$ puts probability 1 on action $T$ from period $k_\la+1$ onwards, by using (\ref{=ep}) for $k_\la$ we obtain 
\begin{eqnarray*}
\bar{g}^\lambda_\infty(\sigma^\star_\la,L[k]) &\leq& \bar{g}^\lambda_\infty(\sigma^\star_\la,L[k_\la+1])\\
 &=& \bar{g}^\lambda_{k_\la}(\sigma^\star_\la,R^\infty)+(1-\la)^{k_\la}\\
 &<& \bar{g}^\lambda_{k_\la}(\sigma^\star_\la,R^\infty)+(1-\la)^{k_\la} z_{\la ,k_\la+1}\\
 &=& \ep,
\end{eqnarray*}
which proves (\ref{eqL[k].2}).\medskip

\noindent\textbf{Step 3: The main strategy $\sigma^\star_\la$ is the best against $R^\infty$.} We argue that, for every $\lambda\in(0,1)$ and every Markov strategy $\sigma$ for player~1 for which $\bar{g}^\lambda_\infty(\sigma,L[k])\leq\ep$ holds for every $k\in\dN$, we have 
\begin{equation}\label{eqstep3}
\bar{g}^\lambda_\infty(\sigma,R^\infty)\leq \bar{g}^\lambda_\infty(\sigma^\star_\la,R^\infty).
\end{equation}\medskip

\noindent\textbf{Proof for step 3: } Let $\lambda\in(0,1)$ and let $\sigma=(p_k,1-p_k)_{k=1}^\infty$ be such a Markov strategy.  Since $\bar{g}^\lambda_\infty(\sigma,R^\infty)\leq \bar{g}^\lambda_{k_\la}(\sigma,R^\infty)$, it suffices to prove 
\begin{equation*}
\bar{g}^\lambda_{k_\la}(\sigma,R^\infty)\leq \bar{g}^\lambda_\infty (\sigma^\star_\la,R^\infty).
\end{equation*}
To prove this inequality, in view of (\ref{eqR}), it suffices in turn to show that for every every $k\in\{1,\ldots,k_\la\}$ 
\begin{equation}\label{notmore}
\bar{g}^\lambda_{k}(\sigma,R^\infty)\leq \ep-\la(1-\la)^{k-1}-(1-\la)^k z_{\la k}.
\end{equation}
We do so by induction on $k$. So first take $k=1$. Since $u_\la(\sigma,L[1])\leq \ep$ by assumption and $u_\la(\sigma,L[1])=p_1$, we have $p_1\leq\ep$. Hence, 
$$\bar{g}^\lambda_{1}(\sigma,R^\infty)=\la(-1+p_1)\leq -\la+\ep\la = \ep-\la-(1-\la) z_{\la 1},$$
where we used that $z_{\la 1}=\ep$. Thus, (\ref{notmore}) holds for $k=1$. Now assume that (\ref{notmore}) holds for some $k$ and that $k+1\leq k_\la$. Since $\bar{g}^\lambda_\infty(\sigma,L[k+1])\leq\ep$ by assumption and
$$\bar{g}^\lambda_\infty(\sigma,L[k+1])=\bar{g}^\lambda_{k}(\sigma,R^\infty)+(1-\la)^k p_{k+1},$$
we have
$$p_{k+1}\leq\frac{\ep-\bar{g}^\lambda_{k}(\sigma,R^\infty)}{(1-\la)^k}.$$
Therefore,
\begin{eqnarray}\label{not more}
\bar{g}^\lambda_{k+1}(\sigma,R^\infty) &=& \bar{g}^\lambda_{k}(\sigma,R^\infty)+\la(1-\la)^k(-1+p_{k+1})\nonumber\\
 &\leq&\bar{g}^\lambda_{k}(\sigma,R^\infty) - \la(1-\la)^k + \ep\la-\la u_{\la k}(\sigma,R^\infty)\nonumber\\
 &=& (1-\la) \bar{g}^\lambda_{k}(\sigma,R^\infty) -\la(1-\la)^k + \ep\la\nonumber\\
 &\leq& (1-\la) [\ep - \la(1-\la)^{k-1}-(1-\la)^k z_{\la k}] -\la(1-\la)^k + \ep\la\nonumber\\
 &=& \ep-\la(1-\la)^k-(1-\la)^{k+1} z_{\la ,k+1},
\end{eqnarray}
we the last equality follows from the definitions of $z_{\la k}$ and $z_{\la ,k+1}$. Thus, (\ref{notmore}) holds for $k+1$ too. The proof for step 3 is now complete.\medskip

\noindent\textbf{Step 4: Conclusion of the proof of the proposition.} We have 
$$\lim_{\la\downarrow 0}\ (1-\la)^{(1-\ep)\tfrac{1-\la}{\la}+1} = e^{-1+\ep},$$
because 
\begin{eqnarray*}
\lim_{\la\downarrow 0}\ \ln \left[ (1-\la)^{(1-\ep)\tfrac{1-\la}{\la}+1} \right] &=& \lim_{\la\downarrow 0}\ \left[ \left( (1-\ep)\tfrac{1-\la}{\la}+1\right) \;\ln (1-\la)\right]\\
 &=& (1-\ep)\; \lim_{\la\downarrow 0}\ \frac{\ln (1-\la)}{\la} \\
 &=& - 1+\ep.
\end{eqnarray*}
Consequently, there is a $\la\star\in(0,1)$ so that for every $\la\in(0,\la\star)$
$$(1-\la)^{(1-\ep)\tfrac{1-\la}{\la}+1}\geq e^{-1}.$$
Let $\la\in(0,\la\star)$. By step 1, it is sufficient to consider a Markov strategy $\sigma$ for player~1. If $\bar{g}^\lambda_\infty(\sigma,L[k])>\ep$ for some $k\in\dN$, then statement of the proposition is valid. So, suppose that $\bar{g}^\lambda_\infty(\sigma,L[k])\leq\ep$ for all $k\in\dN$. Then, by (\ref{eqstep3}) and (\ref{eqR}), we obtain 
$$\bar{g}^\lambda_\infty(\sigma,R^\infty)\leq \bar{g}^\lambda_\infty(\sigma^\star_\la,R^\infty)<\ep-(1-\la)^{k_\la}.$$
Because $z_{\la ,k_\la}\leq 1$, we obtain 
$$k_\la\leq (1-\ep)\tfrac{1-\la}{\la}+1,$$
which implies that 
$$\bar{g}^\lambda_\infty(\sigma,R^\infty)<\ep-(1-\la)^{(1-\ep)\tfrac{1-\la}{\la}+1}\leq \ep - \frac{1}{e}=\frac{1}{2e}-\frac{1}{e}=-\frac{1}{2e}=-\ep.$$
Since $\bar{g}^\lambda_\infty(\sigma,R^\infty)<-\ep$, the proof of the proposition is complete. \qed

\subsection{Conditions \eqref{EQ:Sufficient} and \eqref{EQ:non-Necessary} are not necessary for weak approachability}\label{AppA2}

We revisit the Big-Match game of type II from Example \ref{ex5}. 
\[\begin{tabular}{l|l|c|}
\multicolumn{1}{r}{}
 &  \multicolumn{1}{c}{$L^\star$}
 & \multicolumn{1}{c}{$R$} \\
\cline{2-3}
$T$ & $1$ & $1$ \\
\cline{2-3}
$B$ & $0$ & $-1$ \\
\cline{2-3}
\end{tabular}\]\vspace{0.1cm}

Consider the set $\cC=\{0\}$. As discussed earlier, Conditions \eqref{EQ:Sufficient} and \eqref{EQ:non-Necessary} are not satisfied for $\cC$. Now we argue that player~1 can weakly approach $\cC$.  

The idea (formalized later on) behind the approachability strategy, for the $T$--times repeated game, is the following. At stage $1$, player~1 plays action $T$ with probability $p_1=0$. If the game absorbs at stage 1, the payoff is 0 and is in $\cC$. Otherwise, at stage $2$, player~1 plays action $T$ with probability $p_2=1/(T-1)$. If the game absorbs at stage 2, the total payoff is 0, while otherwise the average payoff up to stage 2 is 
$$\frac{1}{2}(-1-\frac{T-3}{T-1})=-\frac{T-2}{T-1}.$$
By following this idea, the probability of playing $T$ at stage $3$ is then $p_3=2/(T-1)$ and the average payoff up to stage 3 is $-\frac{T-3}{T-1}$, etc.

This ensures that at stage $T$, the cumulative payoff is exactly equal to $0$.

\medskip

This technique can be generalized and formalized as follows. For a mixed action $\bx\in \Delta(\bI)$ for player~1, we use the shorter notations $g_R(\bx)=g(\bx,R)$ and $g^*_L(\bx)=g^*(\bx,L)$. In the remaining part of this subsection, we do not need to assume that the target set $\cC$ is convex.

\begin{proposition}
If a set $\cC$ is weakly approachable, then there is a measurable mapping $\xi : [0,1] \rightarrow \Delta(\bI) $ such that for almost every $t \in [0,1]$, $\int_0^t g_R(\xi (s))ds +(1-t) g^\star_L(\xi (t)) \in \cC$.
\end{proposition}

\textbf{Proof.} Suppose that player~1 can weakly approach $\mathcal{C}$. Then, for each $\varepsilon >0$, there is $T_{\varepsilon}$, such that for every $T \geq T_{\varepsilon}$, there is $\{x^{T, \varepsilon}(k) \in \Delta(\bI), \ \ k=1,...,T\}$, such that for every $s \in [0,1]$:
$$\sum_{k=1}^{\lfloor sT\rfloor} \frac{g_R(x^{T, \varepsilon}(k))}{T} + (1-\frac{\lfloor sT\rfloor}{T})g^\star_L(x^{T, \varepsilon}(\lfloor sT\rfloor+1)) \in \cC+\varepsilon,$$
where $\lfloor r\rfloor$ is the integer part of $r$. Defining $\xi^{T, \varepsilon} (s)=x^{T, \varepsilon}(\lfloor sT\rfloor+1)$, we obtain that for every $x\in[0,1]$:
$$\int_{0}^{x} g_R(\xi^{T, \varepsilon}(s)) ds+ (1-\frac{\lfloor xT\rfloor}{T})g^\star_L(\xi^{T, \varepsilon}(x)) \in \mathcal{C}+\varepsilon$$
We conclude by simply tending $T$ to infinity and $\varepsilon$ to zero. 
\qed

\medskip

Up to a continuity issue, we obtain a converse. 

\begin{proposition}\label{propcontinuoustime}
If there is a continuous mapping $\xi : [0,1] \rightarrow \Delta(\bold{I}) $ such that for every $t \in [0,1]$, $\int_0^t g_R(\xi (s))ds +(1-t) g^\star_L(\xi (t)) \in \mathcal{C}$, then $\mathcal{C}$ is weakly approachable.
\end{proposition}

\textbf{Proof.} For each $\varepsilon>0$, let $T_{\varepsilon}$ be sufficiently large so that for every $T\geq T_{\varepsilon}$ and every $s$ and $t$ in $[0,1]$, if $|s-t| \leq \frac{1}{T}$ then $\|\xi(s)-\xi(t)\|_1\leq \varepsilon$. Now, for each $T$, defining $x^{T}(k)=\xi(\frac{k}{T})$ we obtain a strategy that satisfies, for every $K\in \N$:
$$\sum_{k=1}^{K} \frac{g_R(x^{T}(k))}{T} + (1-\frac{K}{T})g^\star_L(x^{T}(K+1)) \in \mathcal{C}+\varepsilon.$$
%Assume for the moment that player~2  can only use a Markovian strategy (that is independent of the past), which means here the choice of a period $K$ to stop the game (by playing $L$), player~1 $\varepsilon$-approach in this case $\mathcal{C}$. To obtain the same property if player~2 uses a past dependent strategy (stopping times, dependent of player~1's past actions) we can use the same trick as in Vieille's weak approachability and play big blocks of length $L$  i.i.d strategies $\xi(s)$ before switching. By the law of large numbers, on the block $L$, the average payoff if player~2 plays always $R$ is $g_R(\xi(s))$. Second, if $L$ is very small relatively to $N$, the time spend in the block is small and so the absorbing payoff if player $2$ plays $L$ is what it should be.
Thus,  no matter the time $K$ where player~2 plays $L$, the total average payoff will always be $\varepsilon$-close to $\cC$.
\qed

\medskip

The above condition in Proposition \ref{propcontinuoustime} seems not easy to check, as it is merely a rewriting of the approachability objectives in continuous time.  However, it can be helpful in practice. For instance, it allows us to prove (and to find a strategy) that for any $p\geq 1$, player~1 can weakly approach $\{0\}$ in the following game (recall that, as shown in the previous subsection, player~1 cannot weakly approach $\{0\}$ when $p=0$):
\[\begin{tabular}{l|l|c|}
\multicolumn{1}{r}{}
 &  \multicolumn{1}{c}{$L^\ast$}
 & \multicolumn{1}{c}{$R$} \\
\cline{2-3}
$T$ & $1$ & $p$ \\
\cline{2-3}
$B$ & $0$ & $-1$ \\
\cline{2-3}
\end{tabular}\]

To prove our claim, it is sufficient to find a $C^1$ function $\xi : [0,1] \rightarrow [0,1]$ (where $\xi(s)$ is the probability of playing $T$ at time $s$) such that for every $t$: $$\int_0^t (\xi(s)p-(1-\xi(s))ds+(1-t)\xi(t)=0,$$ which is equivalent, by differentiating,  to $\xi(0)=0$ and for every $t$:$$\xi(t)(p+1)-1-\xi(t)+(1-t)\frac{d\xi(t)}{dt}=0,$$or equivalently
$$\xi(t)p-1+(1-t)\frac{d\xi(t)}{dt}=0, \ \ \xi(0)=0.$$
This differential equation has a $C^1$ solution $\xi(t)=\frac{1}{p}(1-(1-t)^p))$ that belongs to $[0,1]$, satisfies $\xi(0)=0$. This has an interpretation in terms of continuous strategy:
$$(1-t)^p\, B+(1-(1-t)^p)(\frac{1}{p}\, T+(1-\frac{1}{p})\, B).$$ In words, this strategy stipulates  that player~1 starts playing $x_0=B$ (the strategy such that $g^\star_L(x_0)=0$) and then, with time, he increases slightly the probability of $T$ until reaching $x_1=\frac{1}{p}\, T+(1-\frac{1}{p})\, B$. Discretization of this continuous time strategy gives, as in the example when $p=1$, weak approachability strategies.

Finally, since this game does not satisfy Condition \eqref{EQ:Sufficient}, we deduce by Proposition \ref{Prop:UnifType2} that $\{0\}$ is not uniformly approachable by player~1. Further, $\{0\}$ is not uniformly excludable by player~2 either, which follows directly from the fact that $\{0\}$ is weakly approachable by player~1. Consequently, this Big-Match game of type II is not uniformly determined.

Observe that the condition in Proposition \ref{propcontinuoustime} can easily extend to the case where player~2 has many quitting actions (and only one non-quitting action). We just need $\int_0^t g_R(\xi (s))ds +(1-t) g^\star_{j^\ast}(\xi (t)) \in \mathcal{C}$ to hold for all  $t$ and all quitting actions $j^\ast \in J^\ast$. If player~2 has more than one non-quitting action, the continuous time condition becomes more complex and is very related to Vieille's approach: one must prove that player~1 can weakly approach the set $\cC$  if he can guarantee zero in a auxiliarry (and non classical) zero-sum differential game $\Gamma$ played between initial time $0$ and terminal time $1$ in which player~1 chooses a trajectory $x(t)$, and player~2 chooses a trajectory $y(t)$ (supported on non-quitting actions), chooses a quitting time $t \in [0,1]$ and a quitting distribution $y^\ast$ (supported on quitting-actions). The payoff of player~1 is the distance of $\int_0^t g(x(s), y(s))ds +(1-t) g^\star(x (t),y^\ast) $ to the set $\cC$. To prove that the game is determined, one should prove that $\Gamma$ has a value. Those are still  open problems.

%Weak determinacy is still an open problem in this class. To fix it, we must prove that the non classical differential game $\Gamma$ has a value.

%To prove it, if one follows the idea in Vieille, one need to prove that some class of differential games (where player~2 has some quitting actions) has always a value. We delegate this question to future investigations.

\section{On uniform approachability in Big-Match games of type~I}
First we revisit Example \ref{ex4}. 

\[\begin{tabular}{l|l|c|}
\multicolumn{1}{r}{}
 &  \multicolumn{1}{c}{$L$}
 & \multicolumn{1}{c}{$R$} \\
\cline{2-3}
$T^\ast$ & $1$ & $0$ \\
\cline{2-3}
$B$ & $0$ & $-1$ \\
\cline{2-3}
\end{tabular}\]\vspace{0.15cm}

In the next proposition we now show that $\{0\}$ is not uniformly approachable. 

\begin{proposition}\label{prop-bmI}
In the above game, for every strategy $\sigma$ for player~1, there is a strategy $\tau$ for player~2 such that $$\liminf_{T \to \infty} \Big| \E_{\sigma,\tau} \frac{1}{T}\sum_{t=1}^Tg(i_t,j_t) \Big|\geq \frac{1}{10},$$ thus $\{0\}$ is not uniformly approachable. 
\end{proposition}

\textbf{Proof.} Take an arbitrary strategy $\sigma$ for player~1. Let $\tau$ be the stationary strategy for player~2 which uses the mixed action $(\tfrac{1}{2},\tfrac{1}{2})$ at every period. Denote by $q^\star$ the probability, with respect to $\sigma$ and $\tau$, that play  absorbs. If $$\limsup_{T \to \infty}  \E_{\sigma,\tau} \frac{1}{T}\sum_{t=1}^Tg(i_t,j_t) <-\tfrac{1}{10}$$ then we are done. So assume that  $\limsup_{T \to \infty}  \E_{\sigma,\tau} \frac{1}{T}\sum_{t=1}^Tg(i_t,j_t) \geq-\tfrac{1}{10}$. Since 
$$ \lim_{T \to \infty} \E_{\sigma,\tau} \frac{1}{T}\sum_{t=1}^Tg(i_t,j_t) =\tfrac{1}{2} q^\star - \tfrac{1}{2}(1-q^\star)=q^\star-\tfrac{1}{2},$$ we have 
$$q^\star\geq -\tfrac{1}{10}+\tfrac{1}{2}=\tfrac{4}{10}.$$
Now let $n\in\N$ be so large that the probability $q_n$ that play absorbs with respect to $\sigma$ and $\tau$ before period $n$ is at least $\tfrac{3}{10}$. Denote by $\tau'$ the Markov strategy for player~2 which uses the mixed action $(\tfrac{1}{2},\tfrac{1}{2})$ at all periods before period $n$ and chooses action $L$ from period $n$ onwards. Then 
$$ \limsup_{T \to \infty} \E_{\sigma,\tau'} \frac{1}{T}\sum_{t=1}^Tg(i_t,j_t)\geq \tfrac{1}{2} q_n \geq \tfrac{3}{20} > \tfrac{1}{10},$$
so the proof is complete.
\qed

\medskip

The Big-Match game of type I above illustrates the complexity of the sufficient condition needed to guarantee uniform approachability. Indeed, let us consider the point of view of player~2. To prove that $\{0\}$ is excludable, we must construct a strategy such that the average payoff is asymptotically $\varepsilon$ away from 0, against all strategy of player~1. Let us denote by $y_1$ the probability of playing $L$ at the first stage. Obviously, $y_1$ must be bigger than $\varepsilon$, otherwise playing $T$ with probability 1 ensures that the asymptotic average payoff is $y_1 \leq \varepsilon$. If we denote by $x_1$ the probability of playing $T$ at the first stage, then in order to approach $[-1,-\varepsilon]\cup[\varepsilon,1]$, player~2 must  be able to approach all the possible sets $[-1,-\frac{\varepsilon+x_1y_1}{1-x_1}]\cup[\frac{\varepsilon-x_1y_1}{1-x_1},1]$.

As a consequence, uniformly approachable set should be defined recursively: player~2 can approach a set if she can approach a certain family of sets, etc, etc. This idea has been observed and explored by Sorin~\cite{Sor85} on $2\times2$ Big-Match games of type I. %Also, in an unpublished note, Sorin~\cite{} proves that in the following game in $\mathbf{R}^2$s%
%
%\[\begin{tabular}{l|l|c|}
%\multicolumn{1}{r}{}
% &  \multicolumn{1}{c}{$L$}
% & \multicolumn{1}{c}{$R$} \\
%\cline{2-3}
%$T$ & $(0,1)^\star$ & $(1,0)^\star$ \\
%\cline{2-3}
%$B$ & $(1,0)$ & $(0,1)$ \\
%\cline{2-3}
%\end{tabular}\]
%
%The set $\cC=\{(x,y): x\geq \frac{3}{8}, y\geq \frac{3}{8} \}$ is neither uniformly approchable nor uniformly excludable. Nevertheless, we will exhibit simple geometric conditions that ensures weak approachability or weak excludability of convex closed sets in all Big Match games of type I.
\section{On almost-sure approachability in Big-Match games}\label{SE:ASApproach}

We consider in this section almost sure approachability. A closed and convex set $\cC \subset \R^d$ is uniformly almost surely approachable if the following condition holds
$$\forall \varepsilon >0, \exists \sigma, \exists T_\varepsilon \in \N, \forall \tau, \P_{\sigma,\tau}\Big\{\exists \, T\geq T_\varepsilon,\  d_\cC\Big( \frac{1}{T}\sum_{t=1}^Tg(i_t,j_t) \Big) \geq \varepsilon\Big\} \leq \varepsilon.
$$
Similarly, $\cC \subset \R^d$ is weakly almost surely approachable if 
$$\forall \varepsilon >0, \exists T_\varepsilon \in \N,  \forall T\geq T_\varepsilon, \exists \sigma_T, \forall \tau, \P_{\sigma_T,\tau}\Big\{ d_\cC\Big( \frac{1}{T}\sum_{t=1}^Tg(i_t,j_t) \Big) \geq \varepsilon\Big\} \leq \varepsilon.
$$
The condition for uniform almost surely approachability can be quite easily obtained.

\begin{proposition}
Let $\cC \subset \R^d$ be a closed and convex set, then in
\begin{description}
\item[BM games of type I:] $\cC$ is uniformly almost surely approachable  if and only if there exists $i^\star \in \cI^\star$ such that $g(i^*,j) \in \cC$ for all $j \in \cJ$ or 
$$
\forall y \in \Delta(\cJ), \exists x \in \Delta(\cI), g(x,y) \in \cC
$$
\item[BM games of type II:] Let $\cI_\cC:=\Big\{ i \in \cI; g(i,j^\star) \in \cC, \ \forall j^\star \in \cJ^\star\Big\}$, then 
$\cC$ is uniformly almost surely approachable  if and only if 
$$
\forall y \in \Delta(\cJ), \exists x \in \Delta(\cI_\cC), g(x,y) \in \cC
$$
\item[Generalized quitting games:] If  $\cI_\cC = \emptyset$  then $\cC$ is uniformly almost surely approachable  if and only if there exists $i^\star \in \cI^\star$ such that $g(i^\star,j) \in \cC$ for all $j \in \bJ$.

If $\cI_\cC \neq \emptyset$, then $\cC$ is uniformly almost surely approachable  if and only if either there exists $i^\star \in \cI^\star$ such that $g(i^\star,j) \in \cC$ for all $j \in \cJ$ or 
$$
\forall y \in \Delta(\cJ), \exists x \in \Delta(\cI_\cC), g(x,y) \in \cC
$$
\end{description}
\end{proposition}
\textbf{Proof.} We consider each case independently.
\begin{description}
\item[BM games of type I:] The fact that the condition is sufficient is immediate: in the first case, player 1 just has to play $i^\star$ at the first stage; in the second case, he just needs to follow the classical Blackwell strategy.

To prove that the condition is necessary, assume that it does not hold. Using Blackwell \cite{Bla56} results (see also Perchet \cite{Per14jfg}), this implies that there exists $y \in \Delta(\cJ)$ with full support such that $g(y):=\{g(x,y)\ ; \ x \in \Delta(\cI)\}$ is $\delta$-away from $\cC$. We denote by $\eta >0$ the smallest probability put on a pure action by $y$. Then Player 2 just has to play i.i.d.\ this mixed action. On the event where the game is not absorbed, then the average payoff converges to the set $g(y)$ thus is $\delta$-away from $\cC$. As a consequence, to approach $\cC$, player 1 must enforce absorption with probability $1-\varepsilon$. But since for every $i^\star \in \cI^\star$ there exists $j\in\cJ$ such that $g(i^\star,j)\not\in \cC$, this implies that with probability at least $(1-\varepsilon)\eta >\eta/2>0$ the payoff does not belong to $\cC$.
 \item[BM games of type II:] In this case again, the fact that the condition is sufficient is immediate. On the contrary, assume that it does not hold, then there exists $y \in\Delta(\cJ)$ with full support (and minimal weight bigger than $\eta>0$) such that $g^\cC(y):=\{g(x,y)\ ; \ x \in \Delta(\cI_\cC)\}$  is $\delta$-away from $\cC$. As a consequence, in order to approach $\cC$, player 1 must use actions that do not belong to $\cI_\cC$ with total probability at least $\delta-\varepsilon$. In particular, it must exist one stage where player 1 uses an action $i^\star \not\in \cI_\cC$ with probability at least $\delta-\varepsilon$; at that stage, player 2 just needs to play the associated action $j^\star \in \cJ^\star$ (so that $g(i^\star,j^\star)\not\in \cC$) to prevent almost sure uniform approachability.
 \item[Generalized quitting games:] In that case, it is actually the sufficient condition which is tricky to prove if $\cI_\cC \neq \emptyset$ and there exists $i^\star \in \cI^\star$ such that $g(i^\star,j) \in \cC$ for all $j \in \cJ$ (the other cases are immediate). In that case, player 1 just needs to play i.i.d.\ any action in $\cI_\cC$ with probability $(1-\varepsilon)$ and the strategy $i^\star \in \cI^\star$ with probability $\varepsilon$. If player 2 uses at any stage $j^\star \in \cJ^\star$, then the game is absorbed in $\cC$ with probability at least $1-\varepsilon$; on the other hand, if he only uses actions in $\cJ$, then the game is absorbed in $\cC$ with probability 1.
  
To prove that the condition is necessary, one just needs to combine the two above arguments.
\end{description}

\qed

We conclude this section by mentioning that those techniques of proof  do not directly provide a necessary and sufficient condition for the weak almost sure approachability case, except in Big-Match of type I. 

Indeed, consider a Big-Match game of type II and assume that the game is absorbed at some stage $t^\star$, and that the absorbing payoff lies outside $\cC$.  Obviously, this implies that the asymptotic average payoff (as $t$ goes to infinity) lies outside $\cC$, thus uniform approachability fails. On the other hand, this unfortunately does not imply that the average payoff at stage $2t^\star$ is outside $\cC$. 

\end{document}